\documentclass[prd,twocolumn,showpacs,preprintnumbers]{revtex4}
\usepackage{amsmath,amsfonts,amssymb,bm}
\usepackage{graphicx} 
\usepackage[hang,normalsize,bf]{subfigure} 

\begin{document}

\title{\Huge \bf Parton transverse momenta  \\
and direct photon production \\
in hadronic collisions \\
at high energies\\}

\author{T. Pietrycki}
\affiliation{Institute of Nuclear Physics\\
PL-31-342 Cracow, Poland\\}

\author{A. Szczurek}
\affiliation{Institute of Nuclear Physics\\
PL-31-342 Cracow, Poland\\}
\affiliation{University of Rzesz\'ow\\
PL-35-959 Rzesz\'ow, Poland\\}


\begin{abstract}
The invariant cross sections for direct photon production in
hadron-hadron collisions are calculated for several initial energies
(SPS, ISR, S$p \bar p$S, RHIC, Tevatron, LHC) including initial parton
transverse momenta within the formalism of unintegrated
parton distributions (UPDF).
Different approaches from the literature are compared and discussed.
A special emphasis is put on the Kimber-Martin-Ryskin (KMR)
distributions and their extension into the soft region.
Sum rules for UPDFs are formulated and discussed in detail.
We find a violation of naive number sum rules for the KMR UPDFs.
An interesting interplay of perturbative (large $k_t^2$)
and nonperturbative (small $k_t^2$) regions of UPDFs
in the production of both soft and hard photons is identified.
The $k_t$-factorization approach with the KMR UPDFs is inconsistent with
the collinear approach at large transverse momenta of photons.
Kwieci\'nski UPDFs provide very good description of all world data,
especially at SPS and ISR energies.
Off-shell effects are discussed and quantified.
Predictions for the CERN LHC are given.
Very forward/backward regions in rapidity at LHC energy are discussed
and a possibility to test unintegrated gluon distributions (UGDF)
is presented.

\end{abstract}


\maketitle

\section{Introduction}

It was realized relatively early that the transverse momenta of initial
(before a hard process) partons may play an important role in order to
understand the distributions of produced direct photons, especially
at small transverse momenta (see e.g.\cite{Owens}). The emitted photon
may be produced directly in a hard process and/or from the fragmentation
process. The latter process involves the parton-to-photon fragmentation
functions which are not very well known. The isolation criterion used
now routinly in the analysis of experimental data helps
to reduce the second component almost completely.

The simplest way to include parton transverse momenta is via Gaussian smearing
\cite{Owens,WW98,AM04}. This phenomenological approach is not completely
justified theoretically. One should remember that there are different
reasons for nonzero transverse momenta of incoming partons. First is purely
nonperturbative, related to the Fermi motion of true hadron
constituents. The transverse momenta related to the internal motion
of hadron constituents are believed to be not too large, definitely
smaller than 1 GeV.
The second is of dynamical nature, related to QCD effects involved
in the evolution of the parton cascades.
The latter effect may be strongly dependent on longitudinal momentum
fraction of the parton taking part in the hard (sub)process.

The unintegrated parton distributions (UPDF) are the basic quantities
that take into account explicitly the parton transverse momenta.
The UPDFs have been studied recently in the context of different
high-energy processes
\cite{Gribov_Levin_Ryskin,KMS97,HKSST1,HKSST2,Mariotto,
LS04,LS06,LZ05,LS05}. These works concentrated
mainly on gluon degrees of fredom which play the dominant role
in many processes at very high energies. At somewhat lower energies
also quark and antiquark degrees of freedom become equally important.
Recently the approach which dynamically includes transverse momenta
of not only gluons but also of quarks and antiquarks was applied 
to direct-photon production \cite{LZ05,KMR_photons}.
In these calculations unintegrated parton
distributions proposed by Kimber-Martin-Ryskin \cite{KMR} were used.
In this approach one assumes that the transverse momenta are generated
only in the last step of the evolution ladder.

Up to now there is no complete agreement how to include evolution
effects into the building blocks of the high-energy processes --
the unintegrated parton distributions. In the present paper 
we shall discuss in detail a few approaches how
to include transverse momenta of the incoming partons in order to calculate
distributions of direct photons. 

\section{Unintegrated parton distributions}

In general, there are no simple relations between unintegrated
and integrated parton distributions.
Some of UPDFs in the literature are obtained based on familiar
collinear distributions, some are obtained by solving evolution
equations, some are just modelled or some are even parametrized.
A brief review of unintegrated gluon distributions (UGDFs) that will
be used also here can be found in Ref.\cite{LS06}.
We shall not repeat all details concerning those UGDFs here.
We shall discuss in more details only approaches which treat
unintegrated quark/antiquark distributions.

In some of the approaches mentioned one imposes the following relation
between the standard collinear distributions and UPDFs:
\begin{equation}
a(x,\mu^2) = \int_0^{\mu^2} f_a(x,k_t^2,\mu^2)
\frac{d k_t^2}{k_t^2}   \; ,
\end{equation}
where $a = xq$ or $a = xg$.

Since familiar collinear distributions satisfy sum rules, one can
define and test analogous sum rules for UPDFs.
We shall discuss this issue in more detail in a separate section. 

Below we shall discuss in detail some of the approaches for UPDFs.
Some other approaches are discussed e.g. in Ref.\cite{LS06}.

\subsection{Gaussian smearing}

Due to its simplicity the Gaussian smearing of initial transverse momenta
is a good reference point for other approaches. It allows to study
phenomenologically the role of transverse momenta in several
high-energy processes.
We define simple unintegrated parton distributions:
\begin{equation}
{\cal F}_{i}^{Gauss}(x,k_t^2,\mu_F^2) = x p_{i}^{coll}(x,\mu_F^2)
\cdot f_{Gauss}(k_t^2) \; ,
\label{Gaussian_UPDFs}
\end{equation}
where $p_{i}^{coll}(x,\mu_F^2)$ are standard collinear (integrated)
parton distribution ($i = g, q, \bar q$) and $f_{Gauss}(k_t^2)$
is a Gaussian two-dimensional function:
\begin{equation}
\begin{split}
f_{Gauss}(k_t^2) = \frac{1}{2 \pi \sigma_0^2}
\exp \left( -k_t^2 / 2 \sigma_0^2 \right) / \pi \; . \\ \;
\label{Gaussian}
\end{split}
\end{equation}
The UPDFs defined by Eq.(\ref{Gaussian_UPDFs}) and (\ref{Gaussian})
are normalized such that:
\begin{equation}
\int {\cal F}_{i}^{Gauss}(x,k_t^2,\mu_F^2) \; d k_t^2 = x
p_{i}^{coll}(x,\mu_F^2) \; .
\label{Gaussian_normalization}
\end{equation}
%

\subsection{KMR distributions}

Kimber, Martin and Ryskin proposed a method to construct
unintegrated parton distributions from the conventional
DGLAP parton distributions \cite{KMR}.
Then
\begin{equation}
\begin{split}
f_a(x,k_t^2,\mu^2) &= T_a(k_t,\mu)\\
& \times \left[  \frac{\alpha_s(k_t^2)}{2\pi}
\int_x^{1-\Delta} \sum_{a'} P_{aa'}(z) \;\;
a'\left(\frac{x}{z},k_t^2 \right) dz \right] \; ,
\end{split}
\label{KMR_1}
\end{equation}
where $P_{aa'}(z)$ are splitting functions
and $a'\left(\frac{x}{z},k_t^2\right)$ are parton densities,
where $a'=\frac{x}{z}g$ or $\frac{x}{z}q$.
Angular-ordering constraint $\Delta = \mu/(\mu + |k_t|)$
regulates the soft gluon singularities.
Recently Lipatov and Zotov \cite{LZ05} used this method to calculate
the direct photon spectra. Technically they did not use 
the original KMR method. Instead they have written
\begin{eqnarray}
&&f_q(x,k_t^2,\mu^2)= T_q(k_t^2,\mu^2)\frac{\alpha_s(k_t^2)}{2\pi}
\int_x^1 dz \nonumber
\\
&\times& \bigg[P_{qq}(z)\frac{x}{z}q\left(\frac{x}{z},
k_t^2\right)\Theta(\Delta-z) + P_{qg}(z)\frac{x}{z}
g\left(\frac{x}{z},k_t^2\right) \bigg] \nonumber \; , \\
\label{LZ_KMR1}
\end{eqnarray}
\begin{eqnarray}
&&f_g(x,k_t^2,\mu^2)= T_g(k_t^2,\mu^2)\frac{\alpha_s(k_t^2)}{2\pi}
\int_x^1 dz \nonumber
\\
&\times& \bigg[\sum_q P_{gq}(z)\frac{x}{z}q\left(\frac{x}{z},
k_t^2\right)+ P_{gg}(z)\frac{x}{z}
g\left(\frac{x}{z},k_t^2\right)\Theta(\Delta-z)  \bigg] \; . \nonumber \\
\label{LZ_KMR2}
\end{eqnarray}
In the following we shall call it LZ KMR prescription. 

The virtual corrections are resummed via Sudakov form factors:
\begin{equation}
{\mathrm ln}\; T_q(k_t^2,\mu^2) = - \int_{k_t^2}^{\mu^2}
\frac{d{\mathrm p}_t^2}{{\mathrm p}_t^2}
\frac{\alpha_s({\mathrm p}_t^2)}{2\pi}
\int_0^{z_{max}}dz \; P_{qq}(z) \; ,
\label{sudakov_quark}
\end{equation}
\begin{widetext}
\begin{equation}
{\mathrm ln}\; T_g(k_t^2,\mu^2) = - \int_{k_t^2}^{\mu^2}
\frac{d{\mathrm p}_t^2}{{\mathrm p}_t^2}
\frac{\alpha_s({\mathrm p}_t^2)}{2\pi}
\left[n_f \; \int_0^1dz \; P_{qg}(z) +
\int_{z_{min}}^{z_{max}}dz \; z \; P_{gg}(z)\right]  \; ,
\label{sudakov_gluon}
\end{equation}
\end{widetext}
where $z_{max} = 1 - z_{min} = \mu/(\mu + |{\mathrm p}_t|)$.

The KMR method presented above can be used for transverse momenta
$k_t^2 > k_{t,0}^2$. In the present paper we assume saturation 
of UPDFs for $k_t^2 < k_{t,0}^2$.
This is a bit arbitrary procedure. We shall discuss the consequences of
the procedure for physical observables.

\subsection{Sum Rules for KMR UPDFs}
In order to gain more insight into the KMR distributions described
shortly in the previous section in the following section we shall 
formulate and check some sum rules.

Let us start from the valence number sum rules.
We define the following integrals for up quarks:
\begin{center}
\begin{equation}
\begin{split}
N_{u_{val}}&(\mu^2,k_{t,0}^2) \equiv
\int_0^1 \frac{dx}{x} \int_0^{\mu^2} d k_t^2 
\\
&\times \left[ f_u^{KMR}(x,k_t^2,\mu^2)
-f_{\bar u}^{KMR}(x,k_t^2,\mu^2)\right] 
\\ \; 
\end{split}
\end{equation}
\end{center}
and for down quarks:
\begin{center}
\begin{equation}
\begin{split}
N_{d_{val}}&(\mu^2,k_{t,0}^2) \equiv
\int_0^1 \frac{dx}{x} \int_0^{\mu^2} d k_t^2 
\\
&\times \left[ f_d^{KMR}(x,k_t^2,\mu^2)
-f_{\bar d}^{KMR}(x,k_t^2,\mu^2)\right] \; .
\\ \;
\end{split}
\end{equation}
\end{center}

\begin{figure}[!h] 
\begin{center}
\includegraphics[width=.4\textwidth]{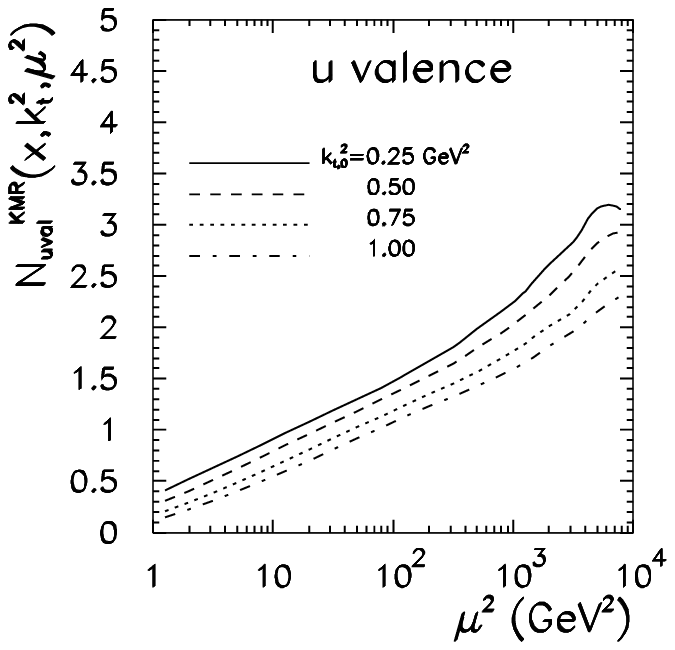}
\includegraphics[width=.4\textwidth]{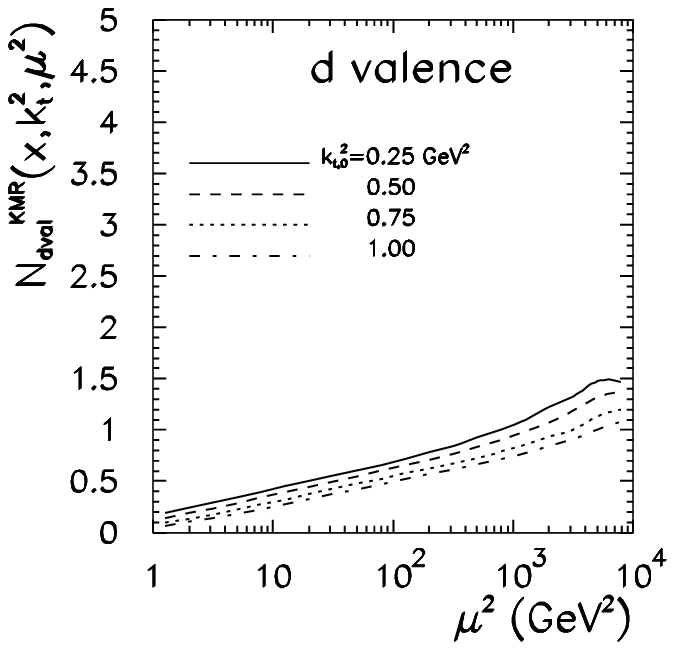}
\caption[*]{
The number sum rule as a function of $\mu^2$
using standard KMR prescription with
Sudakov form factor for up valence quarks (upper panel)
and down valence quarks (lower panel)
for several values of $k_{t,0}^2$.
\label{fig:number_sum_rules}
}
\end{center}
\end{figure}

The parameter $k_{t,0}^2$ is implicit for the KMR distributions as discussed
in the previous section.
Naively one would expect: $N_{u_{val}}$ = 2 and $N_{d_{val}}$ = 1.
We shall check the dependence of these quantities on $\mu^2$
and the freezing parameter $k_{t,0}^2$.
The results are shown in Fig.\ref{fig:number_sum_rules}.

\begin{figure}[!htb] 
\begin{center}
\includegraphics[width=.4\textwidth]{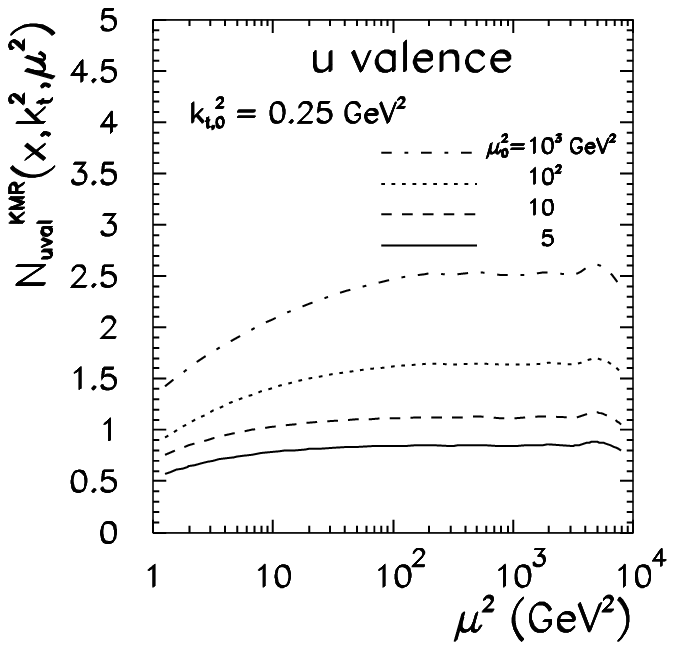}
\includegraphics[width=.4\textwidth]{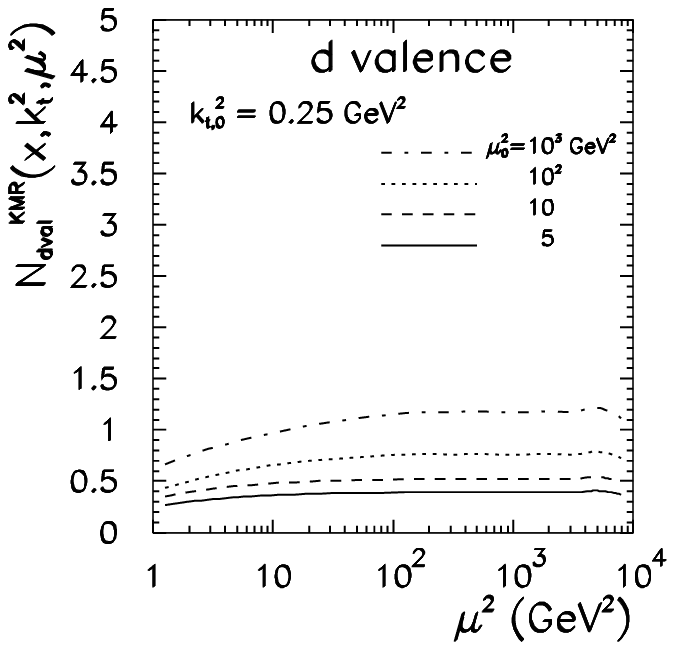}
\caption[*]{
The modified number sum rule as a function of $\mu^2$
using standard KMR prescription with
Sudakov form factor for up valence quarks (upper panel)
and down valence quarks (lower panel)
for several values of $\mu_0^2$.
\label{fig:modified_number_sum_rules}
}
\end{center}
\end{figure}

Somewhat surprisingly the results depend strongly on $\mu^2$ 
and the freezing parameter $k_{t,0}^2$. The results are 
identical for the standard KMR prescription
and the one proposed by Lipatov and Zotov.
In the integrals above the parameter $\mu^2$ occurs as an argument
of the parton distributions as well as the upper limit
of the internal integral.
It seems interesting to allow for independent parameters in the two
places. Therefore we define new quantities for up quarks:
\begin{equation}
\begin{split}
N^0_{u_{val}}&(\mu^2,\mu_0^2,k_{t,0}^2) \equiv
\int_0^1 \frac{dx}{x} \int_0^{\mu^2} d k_t^2
\\
&\times\left[ f_u^{KMR}(x,k_t^2,\mu_0^2)
-f_{\bar u}^{KMR}(x,k_t^2,\mu_0^2)\right] \\ \; 
\end{split}
\end{equation}
and for down quarks:
\begin{center}
\begin{equation}
\begin{split}
N^0_{d_{val}}&(\mu^2,\mu_0^2,k_{t,0}^2) \equiv
\int_0^1 \frac{dx}{x} \int_0^{\mu^2} d k_t^2
\\
&\times\left[ f_d^{KMR}(x,k_t^2,\mu_0^2)
-f_{\bar d}^{KMR}(x,k_t^2,\mu_0^2)\right] \; . \\ \;
\end{split}
\end{equation}
\end{center}
The results for several $\mu_0^2$ are shown in 
Fig.\ref{fig:modified_number_sum_rules}.

Now a saturation of the sum rules for $\mu^2$ larger than
100 GeV$^2$ can be observed.

\begin{figure}[!htb] 
\begin{center}
\includegraphics[width=.4\textwidth]{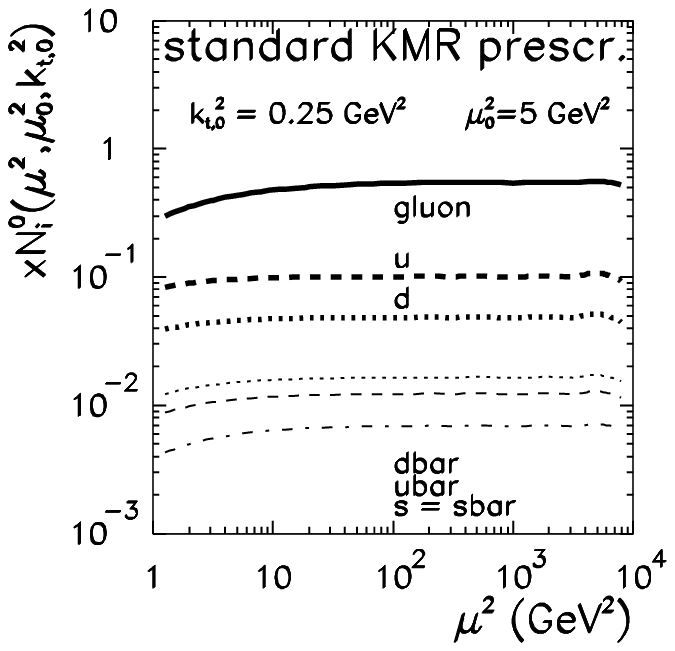}
\includegraphics[width=.4\textwidth]{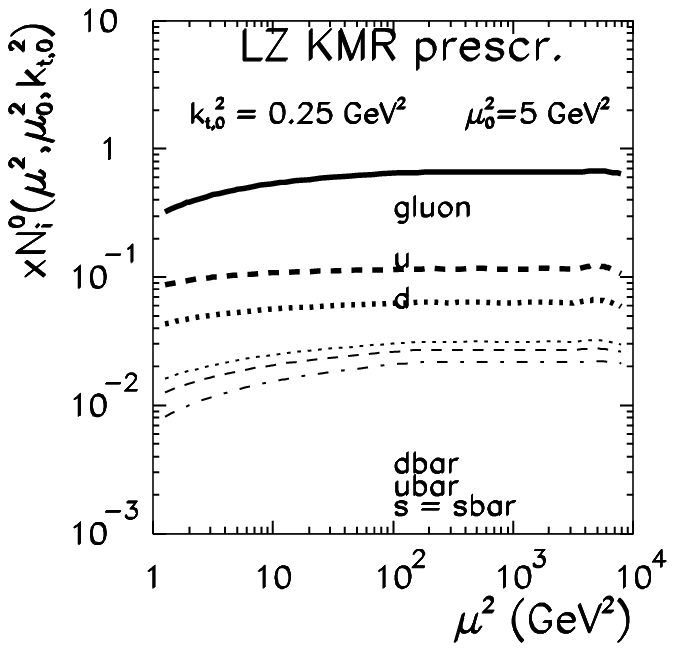}
\caption[*]{
The momentum sum rule as a function of $\mu^2$
using standard KMR prescription (upper panel)
and Lipatov-Zotov prescription (lower panel) with
Sudakov form factor. Here $k_{t,0}^2 = 0.25$ GeV$^2$.
\label{fig:con_momentum_sum_rules}
}
\end{center}
\end{figure}

Another interesting quantity is:
\begin{equation}
xN^0_{i}(\mu^2,\mu_0^2,k_{t,0}^2) \equiv
\int_0^1 dx \int_0^{\mu^2}\left[ f_i^{KMR}(x,k_t^2,\mu_0^2) \right]
 \; d k_t^2
\end{equation}
which can be interpreted as the contribution of parton of a given
type $i$ to the momentum sum rule.
In Fig.\ref{fig:con_momentum_sum_rules} we show contributions
for $g, u, \bar u, d, \bar d, s = \bar s$ as a function of the scale
parameter $\mu^2$.

Again the defined above integrals are functions
of the scale $\mu^2$. In this case there are huge differences
for the standard (left panel) and LZ (right panel) prescription.
These differences cancel as far as valence quarks are considered,
which can be seen by inspection of Eq.(\ref{KMR_1}) 
and/or Eq.(\ref{LZ_KMR1}), Eq.(\ref{LZ_KMR2}).

\subsection{Kwieci\'nski unintegrated parton distributions}

Kwieci\'nski has shown that the evolution equations
for unintegrated parton distributions takes a particularly
simple form in the variable conjugated to the parton transverse momentum.
In the impact-parameter space the Kwieci\'nski equation
takes the following simple form
\begin{eqnarray*}
&&{\partial{\tilde f_{NS}(x,b,\mu^2)}\over \partial \mu^2}
={\alpha_s(\mu^2)\over 2\pi \mu^2}  \int_0^1dz  \, 
P_{qq}(z) 
\\
&&\times \bigg[\Theta(z-x)\,J_0((1-z) \mu b)\,
{\tilde f_{NS}\left({x\over z},b,\mu^2 \right)}
-{\tilde f_{NS}(x,b,\mu^2)} \bigg]  \; , 
\end{eqnarray*}
\vskip 1.0cm
\begin{eqnarray*}
{\partial{\tilde f_{S}(x,b,\mu^2)}\over \partial \mu^2} 
={\alpha_s(\mu^2)\over 2\pi \mu^2} \int_0^1 dz
\bigg\{\Theta(z-x)\,J_0((1-z) \mu b)&&
\\
\times \bigg[P_{qq}(z)\,
{\tilde f_{S}\left({x\over z},b,\mu^2 \right)}
+ P_{qg}(z)\, {\tilde f_{G}\left({x\over z},b,\mu^2 \right)}\bigg]&&
\\
- \bigg[zP_{qq}(z)+zP_{gq}(z)\bigg]\,
{\tilde f_{S}(x,b,\mu^2)}\bigg\}  \; , &&
\end{eqnarray*}
\begin{eqnarray}
{ \partial {\tilde f_{G}(x,b,\mu^2)}\over \partial \mu^2}
={\alpha_s(\mu^2)\over 2\pi \mu^2} \int_0^1 dz
\bigg\{\Theta(z-x)\,J_0((1-z) \mu b) && \nonumber
\\
\times \bigg[P_{gq}(z)\,
{\tilde f_{S}\left({x\over z},b,\mu^2 \right)}
+ P_{gg}(z)\, {\tilde f_{G}\left({x\over z},b,\mu^2 \right)}\bigg]&&
\nonumber
\\
-\bigg[zP_{gg}(z)+zP_{qg}(z)\bigg]\, 
{\tilde f_{G}(x,b,\mu^2)}\bigg\} \; . && \nonumber
\\
\label{kwiecinski_equations}
\end{eqnarray}
We have introduced here the short-hand notation
\begin{equation}
\begin{split}
\tilde f_{NS}&= \tilde f_u - \tilde f_{\bar u}, \;\;
                 \tilde f_d - \tilde f_{\bar d} \; ,  \\
\tilde f_{S}&= \tilde f_u + \tilde f_{\bar u} + 
                \tilde f_d + \tilde f_{\bar d} + 
                \tilde f_s + \tilde f_{\bar s} \; . 
\end{split}
\label{singlet_nonsinglet}
\end{equation}
The unintegrated parton distributions in the impact factor
representation are related to the familiar collinear distributions
as follows
\begin{equation}
\tilde f_{k}(x,b=0,\mu^2)=\frac{x}{2} p_k(x,\mu^2) \; .
\label{uPDF_coll_1}
\end{equation}
On the other hand, the transverse momentum dependent UPDFs are related
to the integrated parton distributions as
\begin{equation}
x p_k(x,\mu^2) =
\int_0^{\infty} d k_t^2 \; f_k(x,k_t^2,\mu^2) \; .
\label{uPDF_coll_2}
\end{equation}
The two possible representations, in the momentum space and in the 
impact parameter space, are interrelated via Fourier-Bessel
transform
\begin{equation}
  \begin{split}
    &{f_k(x,k_t^2,\mu^2)} =
    \int_{0}^{\infty} db \;  b J_0(k_t b)
    {{\tilde f}_k(x,b,\mu^2)} \; ,
    \\
    &{{\tilde f}_k(x,b,\mu^2)} =
    \int_{0}^{\infty} d k_t \;  k_t J_0(k_t b)
    {f_k(x,k_t^2,\mu^2)} \; .
  \end{split}
\label{Fourier}
\end{equation}
The index k above numerates either gluons (k=0), quarks (k$>$ 0) or
antiquarks (k$<$ 0).
While physically $f_k(x,k_t^2,\mu^2)$ should be positive,
there is no obvious reason for such a limitation for
$\tilde f_k(x,b,\mu^2)$.

In the following we use leading-order parton distributions
from Ref.\cite{GRV98} as the initial condition for QCD evolution.
The set of integro-differential equations in b-space
was solved by the method based on the discretisation made with
the help of the Chebyshev polynomials (see \cite{Kwiecinski}).
Then the unintegrated parton distributions were put on a grid
in $x$, $b$ and $\mu^2$ and the grid was used in practical
applications for Chebyshev interpolation. 

For the calculation of inclusive and coincidence cross section
for the photon production (see next section) the parton distributions
in momentum space are more useful.
These calculation requires a time-consuming multi-dimensional
integrations. Therefore an explicit calculation of the momentum-space 
of the Kwieci\'nski UPDFs
via Fourier transform for needed in the main calculation values of
$(x_1,k_{1,t}^2)$ and $(x_2,k_{2,t}^2)$ (see next section)
is not possible.
Therefore it becomes a neccessity to prepare auxiliary grids of
the momentum-representation UPDFs before
the actual calculation of the cross sections.
These grids are then used via a two-dimensional interpolation
in the spaces $(x_1,k_{1,t}^2)$ and $(x_2,k_{2,t}^2)$
associated with each of the two incoming partons.

\begin{figure}[!h] 
\begin{center}
\includegraphics[width=.4\textwidth]{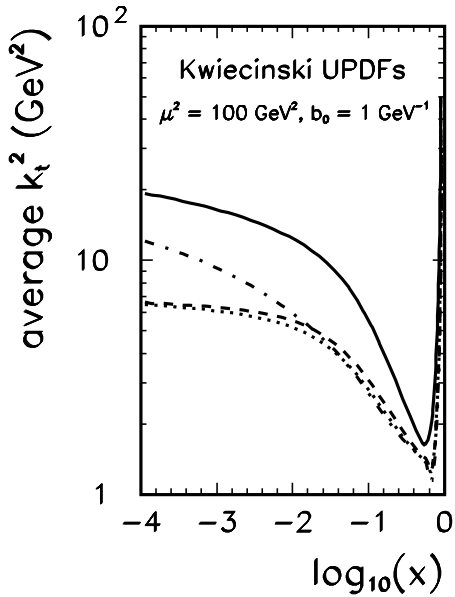}
\caption[*]{
$< k_t^2 >$ as a function of $x$ for different unintegrated
parton distributions: solid - glue, dashed - $u_{val}$, dotted -
$d_{val}$ and dash-dotted - sea.
\label{fig:ave_kt2_x}
}
\end{center}
\end{figure}

The evolution of the parton cascade leads to a spread of
the transverse momentum of the parton at the end of the cascade
(the parton participating in a hard process).
Let us define the following measure of the spread:
\begin{equation}
< k_t^2 >_{f_k}(x,\mu^2) \equiv 
\frac{\int d k_t^2 \; k_t^2 \; f_k(x,k_t^2,\mu^2)}
     {\int d k_t^2 \;          f_k(x,k_t^2,\mu^2)} \; .
\end{equation}
Above $f_k$ can be either gluon ($k=0$) or quark ($k>0$) 
or antiquark ($k<0$) distribution.
As an example in Fig.\ref{fig:ave_kt2_x} we show
the spread, obtained for different parton species,
as a function of parton longitudinal momentum fraction.

In this calculation the factorization scale was fixed at
$\mu^2 =$ 100 GeV$^2$. 
The Kwieci\'nski evolution leads to increasing spread
with decreasing longitudinal momentum fraction.
The spread for different species of partons is quite different.
In the region of small $x$ the spread in $k_t^2$ for gluons is bigger
than a similar spread for sea and valence quarks.
This is very different than a corresponding spread for Gaussian
distributions which is usually taken to be independent of $x$
and parton species. In addition, the spread of $k_t^2$ for small
values of $x$ is considerably larger than the nonperturbative
spread of the initial Gaussian distributions, taken here identical for
all species and encoded in the model parameter $b_0$.

In contrast to the KMR unintegrated valence quark distributions
the Kwieci\'nski valence quark distributions fulfill the number sum
rules for up:
\begin{equation}
\begin{split}
N_{u_{val}}(\mu^2) &=
\int_0^1 \frac{dx}{x} \int_0^{\infty} \; d k_t^2
\\
& \times \left[ f_u^{Kwiec}(x,k_t^2,\mu^2)
-f_{\bar u}^{Kwiec}(x,k_t^2,\mu^2)\right]  = 2 \\ \; 
\end{split}
\end{equation}
and for down quarks:
\begin{equation}
\begin{split}
N_{d_{val}}(\mu^2) &=
\int_0^1 \frac{dx}{x} \int_0^{\infty}\; d k_t^2 
\\
&\times\left[ f_d^{Kwiec}(x,k_t^2,\mu^2)
-f_{\bar d}^{Kwiec}(x,k_t^2,\mu^2)\right] = 1 \; . 
\end{split}
\end{equation}

\subsection{Mixed distributions}

The calculation with the Kwieci\'nski distributions discussed 
in the previous section shows
that the spread in $k_t^2$ for gluons can be much bigger than
that for quarks/antiquarks. On the other hand in the region of small $x$
there are several
unintegrated gluon distributions available in the literature.
At high energy (small $x$) the contribution of $qg$, $gq$ subprocesses is
larger than the contribution of $q\bar q$, $\bar q q$ subprocesses.
Therefore it seems reasonable to use the different UGDFs from
the literature together with the Gaussian distributions for quarks
and antiquarks as discribed above. Such an approach is especially
justified at forward ($y_{\gamma} \gg$ 0) and
backward photon rapidities ($y_{\gamma} \ll$ 0),
where $x_1 \ll$ 1, $x_2 \sim$ 1
($\langle k_{1,t} \rangle_{glue} > \langle k_{2,t}
\rangle_{q,\bar q}$) or
$x_1 \sim$ 1, $x_2 \ll$ 1
($\langle k_{1,t} \rangle_{q \bar q} < \langle k_{2,t}
\rangle_{glue}$).
\\
\\

\section{UPDFs and photon production}

\begin{figure}[!htb] 
\begin{center}
\includegraphics[width=.45\textwidth]{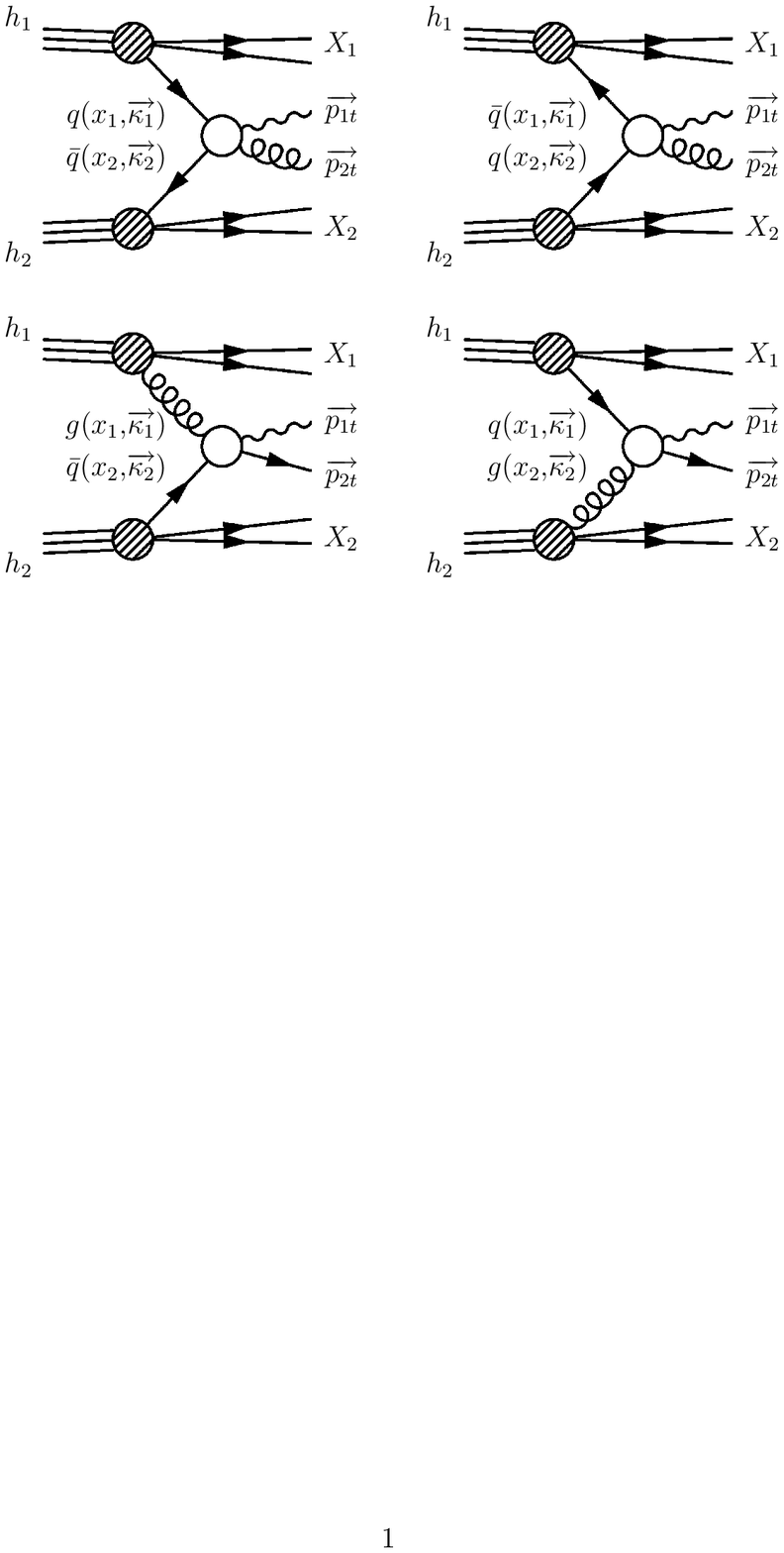}
\caption[*]{
The diagrams included in our $k_t$-factorization approach with
the notation of kinematical variables.
\label{fig:diagrams}
}
\end{center}
\end{figure}

The cross section for the production of a photon and an associated parton (jet)
can be written as
\begin{widetext}
\begin{eqnarray}
\frac{d\sigma(h_1 h_2 \rightarrow \gamma, parton)}
{d^2p_{1,t}d^2p_{2,t}} &=& \int dy_1 dy_2
\frac{d^2 k_{1,t}}{\pi}\frac{d^2 k_{2,t}}{\pi}
\frac{1}{16\pi^2(x_1x_2s)^2} \; \sum_{i,j,k}
\overline{|\mathcal{M}(i j \rightarrow \gamma k)|^2}
\nonumber \\
&\cdot&\delta^2(\vec{k}_{1,t}
+\vec{k}_{2,t}
-\vec{p}_{1,t}
-\vec{p}_{2,t})
f_i(x_1,k_{1,t}^2)f_j(x_2,k_{2,t}^2) \; ,
\label{basic_formula}
\end{eqnarray}
\end{widetext}
where
$\vec{k}_{1,t}$ and $\vec{k}_{2,t}$ are transverse
momenta of incoming partons.
The longitudinal momentum fractions are calculated as
\begin{equation}
x_1 = \frac{m_{1t}}{\sqrt{s}}\mathrm{e}^{-y_1} 
    + \frac{m_{2t}}{\sqrt{s}}\mathrm{e}^{-y_2} \; ,
\end{equation}
\begin{equation}
x_2 = \frac{m_{1t}}{\sqrt{s}}\mathrm{e}^{y_1} 
    + \frac{m_{2t}}{\sqrt{s}}\mathrm{e}^{y_2} \; ,
\end{equation}
where $m_{1t}$ and $m_{2t}$ are respective transverse masses.
In the leading-order approximation: $(i,j,k) = (q,{\bar q},g),
({\bar q},q,g), (q, g, q), (g, q, q)$, etc. (see Fig.\ref{fig:diagrams}).

If one makes the following replacements
\begin{equation}
f_i(x_1,k_{1,t}^2) \rightarrow x_1p_i(x_1)
\delta(k_{1,t}^2)
\end{equation}
and
\begin{equation}
f_j(x_2,k_{2,t}^2) \rightarrow x_2p_j(x_2)
\delta(k_{2,t}^2)
\end{equation}
then one recovers the standard collinear formula (see e.g.\cite{Owens}).

The inclusive invariant cross section for direct photon production can be
written
\begin{widetext}
\begin{equation}
\begin{split}
\frac{d \sigma(h_1h_2 \to \gamma)}{dy_1 d^2p_{1,t}} 
&= \int dy_2 \frac{d^2 k_{1,t}}{\pi} \frac{d^2 k_{2,t}}{\pi}
\left( ... \right) |
_{\vec{p}_{2,t} = \vec{k}_{1,t} + \vec{k}_{2,t} -
  \vec{p}_{1,t}}
\\  
&= \int d k_{1,t} d k_{2,t} \; 
I(k_{1,t},k_{2,t};y_1, p_{1,t}) \; .
\label{inclusive1}
\end{split}
\end{equation}
\end{widetext}
and analogously the cross section for the associated parton (jet) can be
written 
\begin{equation}
\frac{d \sigma(h_1h_2 \to k)}{dy_2 d^2p_{2,t}} =
\int dy_1 \frac{d^2 k_{1,t}}{\pi} \frac{d^2 k_{2,t}}{\pi}
\left( ... \right) |
_{\vec{p}_{1,t} = \vec{k}_{1,t} + \vec{k}_{2,t} -
  \vec{p}_{2,t}} \;.
\label{inclusive2}
\end{equation}
The integrand $I(k_{1,t},k_{2,t};y_1,p_{1,t})$ defined
in Eq.(\ref{inclusive1}) depends strongly on UPDFs used.
We shall return to this interesting issue.

Let us return to the coincidence cross section.
The integration with the Dirac delta function in (\ref{basic_formula})
\begin{equation}
\int dy_1 dy_2
\frac{d^2 k_{1,t}}{\pi}\frac{d^2 k_{2,t}}{\pi}
\left(...\right) \delta^2(...)
\end{equation}
can be performed by introducing the following new auxiliary variables:
\begin{eqnarray}
\vec{Q}_t &=&
\vec{k}_{1,t}+\vec{k}_{2,t} \; ,
\nonumber \\
\vec{q}_t &=&
\vec{k}_{1,t}-\vec{k}_{2,t} \; .
\end{eqnarray}
The jacobian of this transformation is:
\begin{equation}
\frac{\partial 
(\vec{Q}_t,\vec{q}_t)}
{\partial
(\vec{k}_{1,t},
\vec{k}_{2,t})}=
\begin{pmatrix}
1 & 1 \\
1 &-1 \\
\end{pmatrix}
\cdot
\begin{pmatrix}
1 & 1 \\
1 &-1 \\
\end{pmatrix}
=2 \cdot 2 = 4 \; .
\end{equation}
Then our initial cross section can be written as:
\begin{widetext}
\begin{equation}
\begin{split}
\frac{d\sigma(h_1 h_2 \rightarrow \gamma,parton)}
{d^2p_{1,t}d^2p_{2,t}}
&= \frac{1}{4} \int dy_1 dy_2
\;d^2Q_t d^2q_t \;\;\; ( ... ) \;
\delta^2(\vec{Q}_t-\vec{p}_{1,t}
-\vec{p}_{2,t})
\\
&= \frac{1}{4} \int dy_1 dy_2\;\;\;\;\;\;\; 
\underbrace{d^2q_t}\;\;\;\;\; \left(...\right) \;
|_{\vec{Q}_t=\vec{P}_t} 
\\
&= \frac{1}{4} \int dy_1 dy_2\;\;\;\; 
\overbrace{\underbrace{q_tdq_t} \;d\varphi} \; \left(...\right) \; 
|_{\vec{Q}_t=\vec{P}_t}
\\
&= \frac{1}{4} \int dy_1 dy_2\;
\overbrace{\frac{1}{2}dq_t^2 \;d\varphi} \;\;\;\; \left(...\right) \;
|_{\vec{Q}_t=\vec{P}_t} \;.
\end{split}
\end{equation}
\end{widetext}
Above $\vec{P}_t = \vec{p}_{1,t} + \vec{p}_{2,t}$.
Different representations of the cross section are
possible. If one is interested in the distribution of the sum of
transverse momenta of the outgoing particles (parton and photon), 
then it is convenient to write 
\begin{eqnarray}
d^2p_{1,t}\; d^2p_{2,t} &=& \frac{1}{4}d^2P_td^2p_t = 
\frac{1}{4}d\varphi_+P_tdP_t \;d\varphi_-p_tdp_t
\nonumber \\
&=&\frac{1}{4}\;2\pi P_tdP_t \;d\varphi_-p_tdp_t \; .
\end{eqnarray}
If one is interested in studying a two-dimensional map
$p_{1,t} \times p_{2,t}$ then the differential volume element
can be written
\begin{equation}
d^2p_{1,t}\; d^2p_{2,t} = 
d\phi_1 \; p_{1,t} dp_{1,t} \; 
d\phi_2 \; p_{2,t} dp_{2,t} \; .
\end{equation}
Then the two-dimensional map can be written as
\begin{equation}
\frac{d \sigma(p_{1,t},p_{2,t})}{d p_{1,t} d p_{2,t}}
= \int d \phi_1 d \phi_2 p_{1,t} p_{2,t} 
\int d y_1 d y_2 \frac{1}{4} q_t d q_t d\phi_{q_t}\left(...\right).
\label{2-dim-map_a}
\end{equation}
The integrals over $\phi_1$ and $\phi_2$ must be the most external ones.
The integral above is formally a 6-dimensional one.
It is convenient to make the following transformation of variables
\begin{equation}
(\phi_1, \phi_2) \to \left( \phi_{sum} = \phi_1 + \phi_2, 
\; \phi_{dif} = \phi_1 - \phi_2 \right) \; ,
\end{equation}
Explicit formulae for the basic matrix elements are given in Appendix B.

\section{Inclusive photon spectra}

\subsection{Integrands of the inclusive cross sections}

Before we go to the discussion of the dependence
of the invariant cross sections on the values of rapidity
and photon transverse momentum
let us consider the integrand $I(k_{1,t},k_{2,t};y_1,p_{1,t})$
(before integration over $k_{1,t}$ and $k_{2,t}$) 
in Eq.(\ref{inclusive1}).

\begin{figure}[!htb] 
\begin{center}
\begin{tabular}{cc}
\includegraphics[width=.23\textwidth]{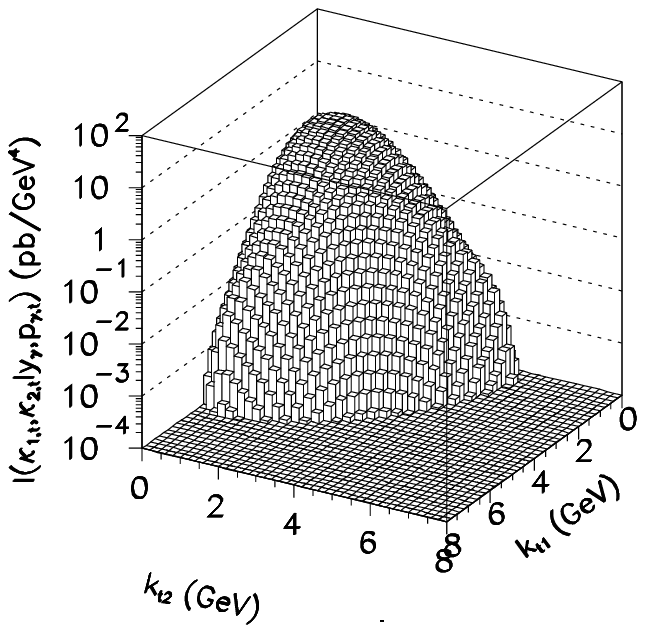}&
\includegraphics[width=.23\textwidth]{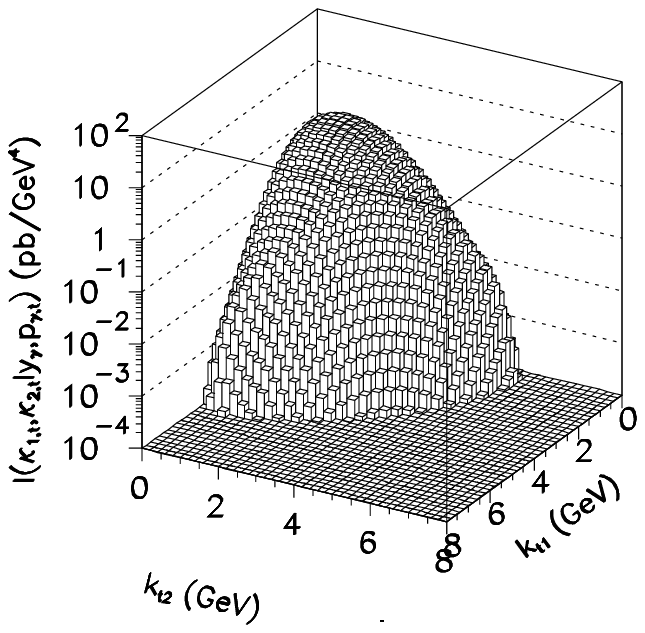}\\
\includegraphics[width=.23\textwidth]{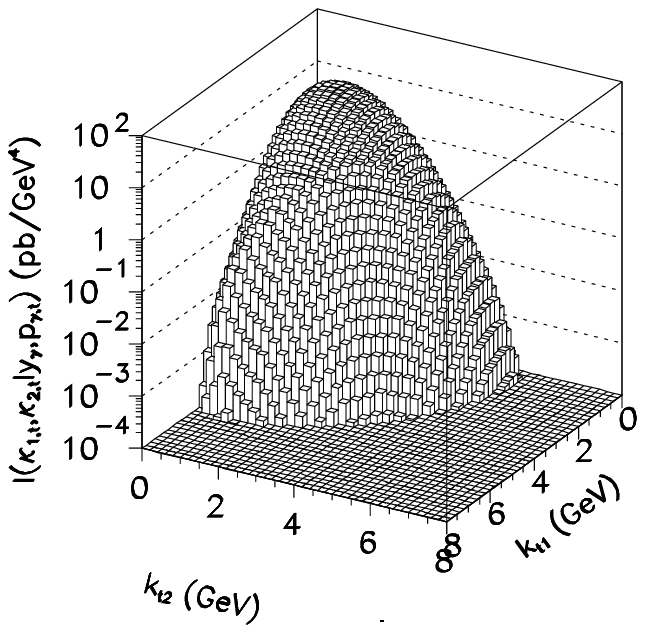}&
\includegraphics[width=.23\textwidth]{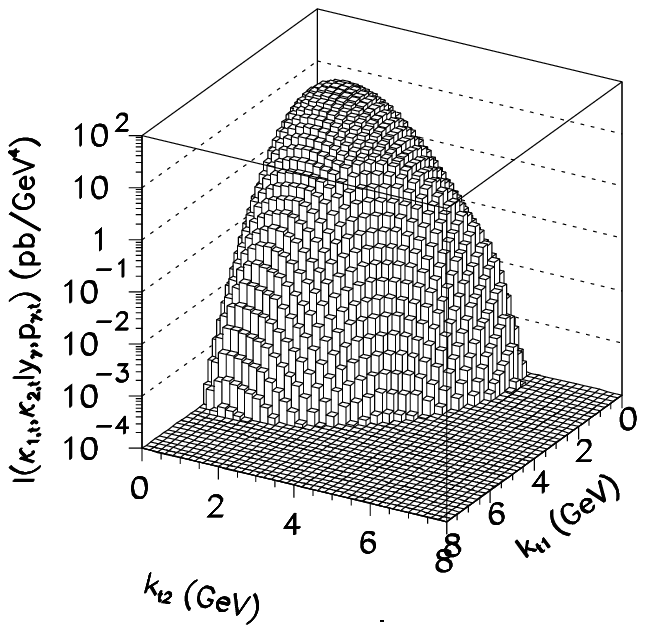}\\
\end{tabular}
\caption[*]{
Integrand under the $k_{1,t}$ and $k_{2,t}$ integration
in the invariant cross section formula for
proton-proton scattering at
$\sqrt{s} = 63$ GeV and y= 0, $p_{t,\gamma}$ = 5 GeV
and Gaussian UPDFs ($\sigma_0$ = 1 GeV).
Off-shell matrix elements for gluons are used.
(a) $q \bar q$, (b) $\bar qq$, (c) $gq$, (d) $qg$.
\label{fig:gauss_kt1_kt2_w63}
}
\end{center}
\end{figure}

In Fig.\ref{fig:gauss_kt1_kt2_w63} we show an example for
$\sqrt{s}$ = 63 GeV, $y$ = 0 and $p_{t,\gamma}$ = 5 GeV.
In this calculation the unintegrated parton distributions based on
GRV collinear parton distributions \cite{GRV95} and Gaussian smearing
($\sigma_0$ = 1 GeV) in parton transverse momenta were used.
This is a rather standard method to ``improve'' the
collinear approach. We do not need to mention that this,
although having some physical motivation,
is a rather ad hoc procedure.
The two-dimensional distributions are peaked for small values
of $k_{1,t}$ and $k_{2,t}$. How fast the distribution
decreases with $k_{1,t}$ and/or $k_{2,t}$ depends on the value 
of the smearing parameter $\sigma_0$.
The larger $\sigma_0$ the slower the decrease.

\begin{figure}[!htb] 
\begin{center}
\includegraphics[width=.23\textwidth]{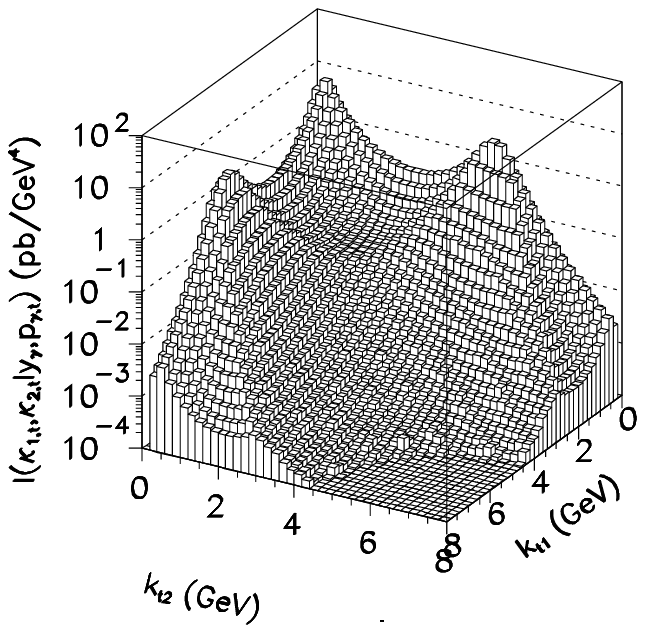}
\includegraphics[width=.23\textwidth]{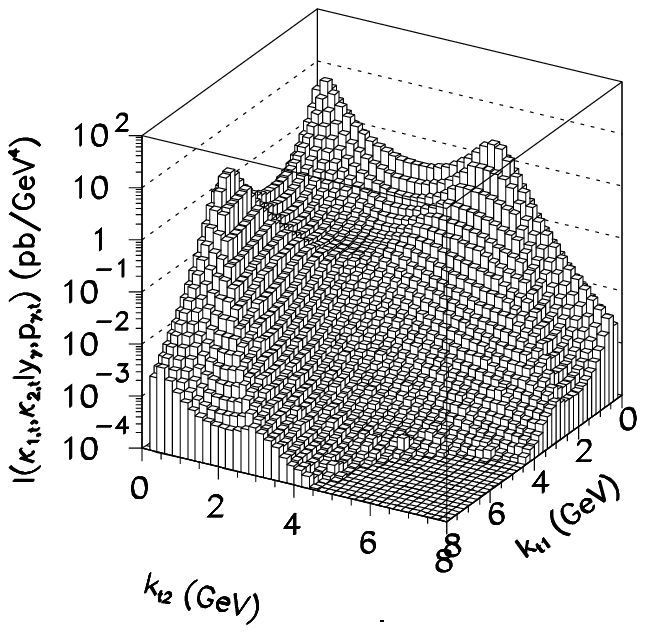}
\includegraphics[width=.23\textwidth]{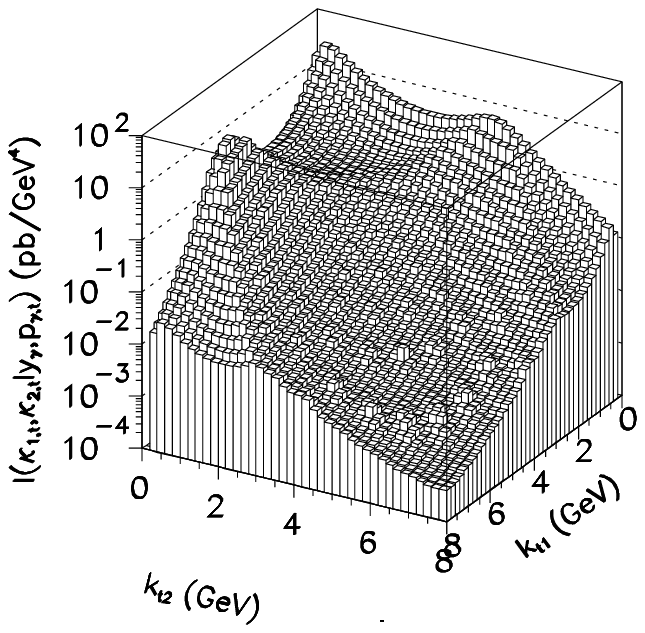}
\includegraphics[width=.23\textwidth]{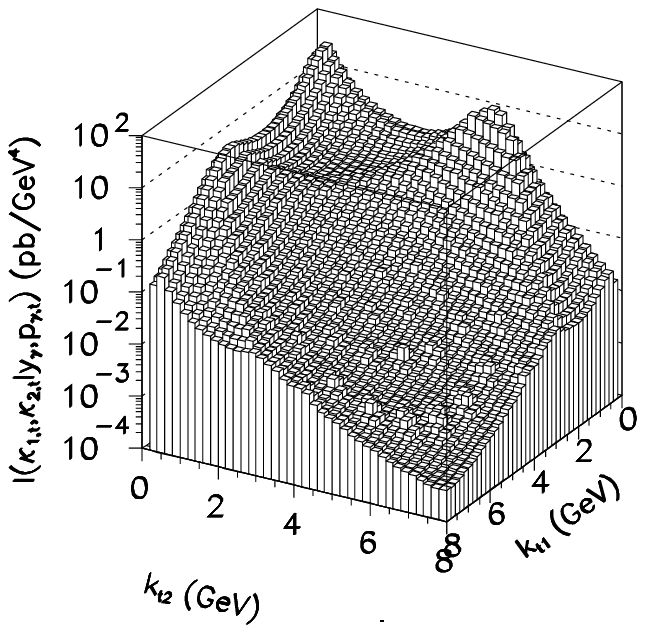}
\caption[*]{
Integrand under the $k_{1,t}$ and $k_{2,t}$ integration
in the invariant cross section formula for
proton-proton scattering at
$\sqrt{s} = 63$ GeV and y= 0, $p_{t,\gamma}$ = 5 GeV
and KMR UPDFs ($k_{t,0}^2$ = 0.25 GeV$^2$).
Off-shell matrix elements for gluons are used.
(a) $q \bar q$, (b) $\bar qq$, (c) $gq$, (d) $qg$.
\label{fig:kmr_kt1_kt2_w63}
}
\end{center}
\end{figure}

\begin{figure}[!htb] 
\begin{center}
\includegraphics[width=.23\textwidth]{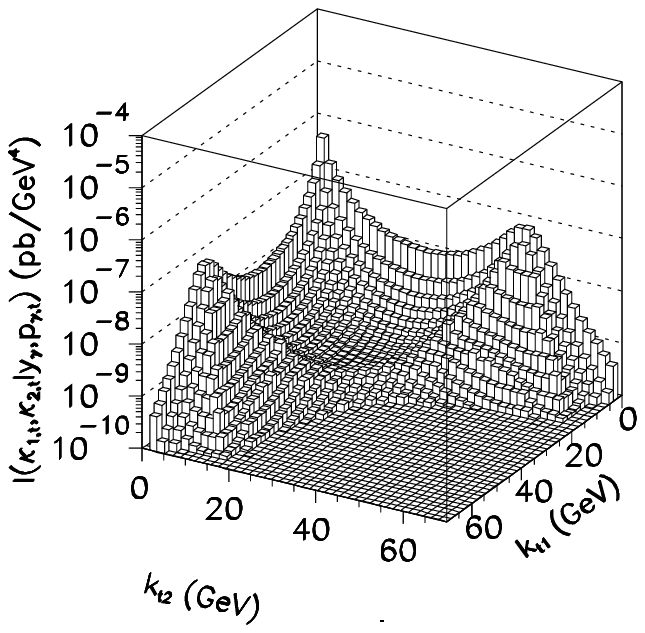}
\includegraphics[width=.23\textwidth]{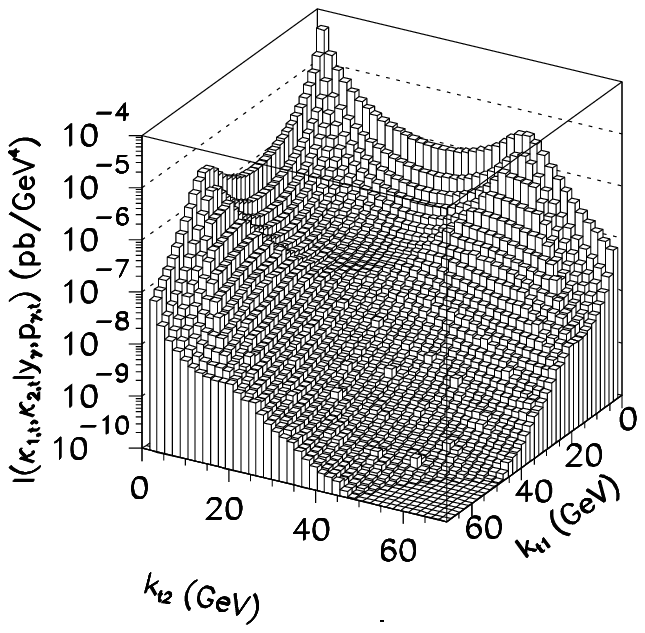}
\includegraphics[width=.23\textwidth]{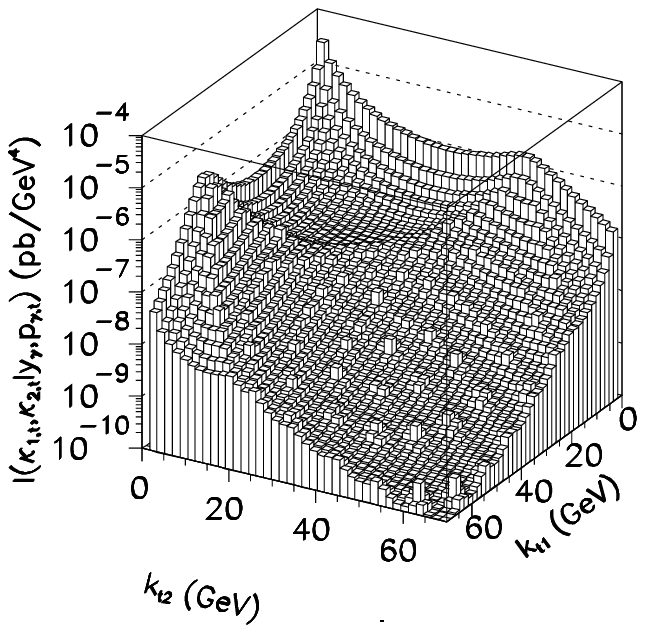}
\includegraphics[width=.23\textwidth]{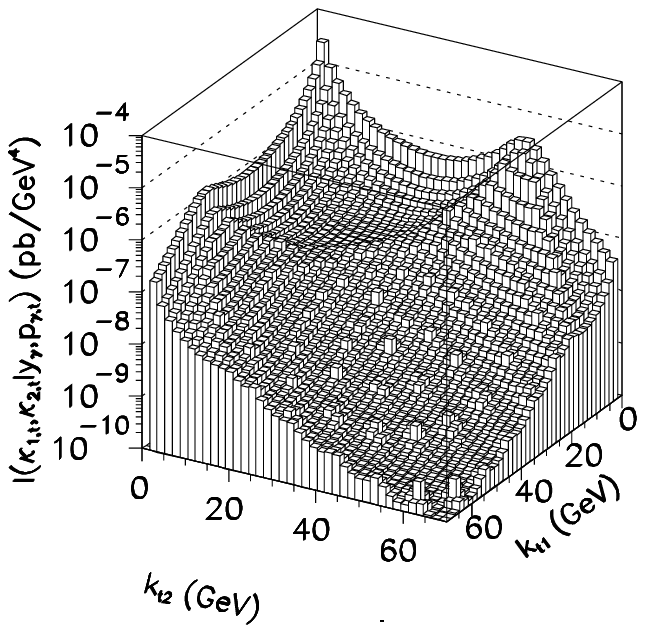}
\caption[*]{
Integrand under the $k_{1,t}$ and $k_{2,t}$ integration
in the invariant cross section formula for proton-antiproton
scattering at $\sqrt{s} = 630$ GeV and y= 0, $p_{t,\gamma}$ = 50 GeV
and KMR UPDFs ($k_{t,0}^2$ = 0.25 GeV$^2$).
Off-shell matrix elements for gluons are used.
(a) $q \bar q$, (b) $\bar qq$, (c) $gq$, (d) $qg$.
\label{fig:kmr_kt1_kt2_w630}
}
\end{center}
\end{figure}

In Fig.\ref{fig:kmr_kt1_kt2_w63} we show 
similar maps ($\sqrt{s}$ = 63 GeV, $y$ = 0, $p_{t,\gamma}$ = 5 GeV)
for the KMR UPDFs.
Three local maxima can be seen in the figure.
A first maximum occurs when both $k_{1,t}$ and $k_{2,t}$
are very small. This is caused by the structure of UPDFs themselves.
The two other maxima occur when $k_{1,t} = p_t$ and $k_{2,t}$
is small or $k_{2,t} = p_t$ and $k_{1,t}$ is small.
These are caused by the structure of matrix elements.
The presence of long tails in $k_t$ in the KMR distributions
is a necessary condition to produce the second and third maxima. 
When $p_t$ increases the second and third maxima move towards
larger $k_{1,t}$ and/or $k_{2,t}$.
This clearly shows that the range of integration must depend
on the value of photon transverse momentum. 
In Fig.\ref{fig:kmr_kt1_kt2_w630} we show some integrand (KMR UPDFs)
but for larger energy $W=630$ GeV and larger photon transverse momentum
$p_{t,\gamma}=50$ GeV.

Fig.\ref{fig:kmr_kt1_kt2_w630} looks 
very much the same as Fig.\ref{fig:kmr_kt1_kt2_w63} 
if the transverse momenta of incoming 
partons are rescaled by the ratio of photon transverse momenta.

\begin{figure}[!htb] 
\begin{center}
\includegraphics[width=.23\textwidth]{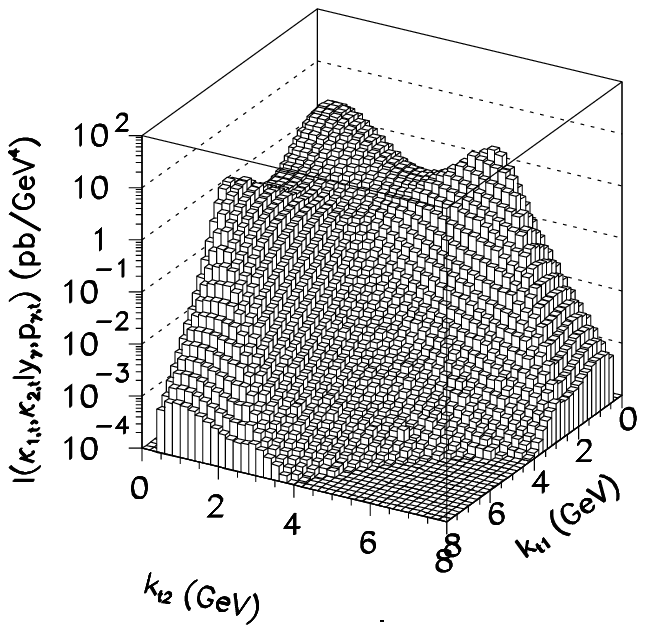}
\includegraphics[width=.23\textwidth]{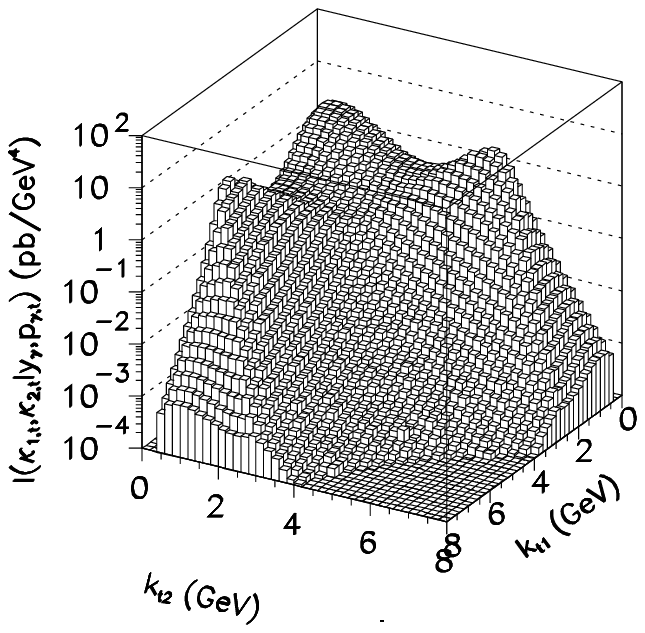}
\includegraphics[width=.23\textwidth]{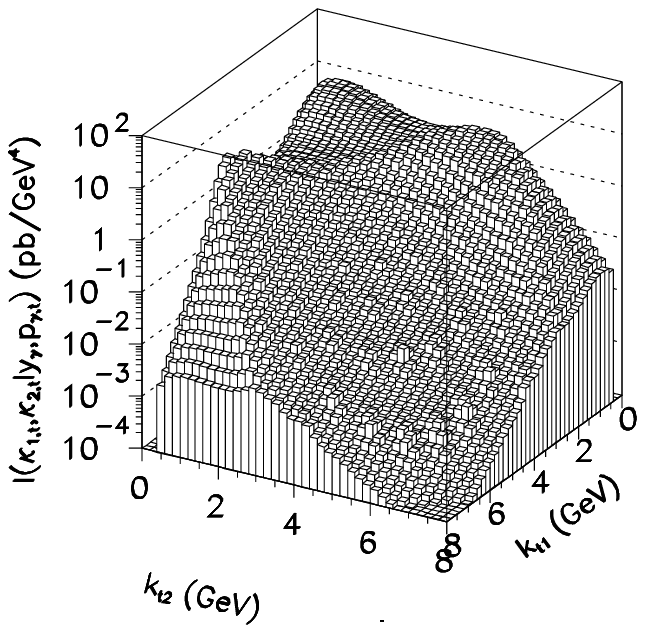}
\includegraphics[width=.23\textwidth]{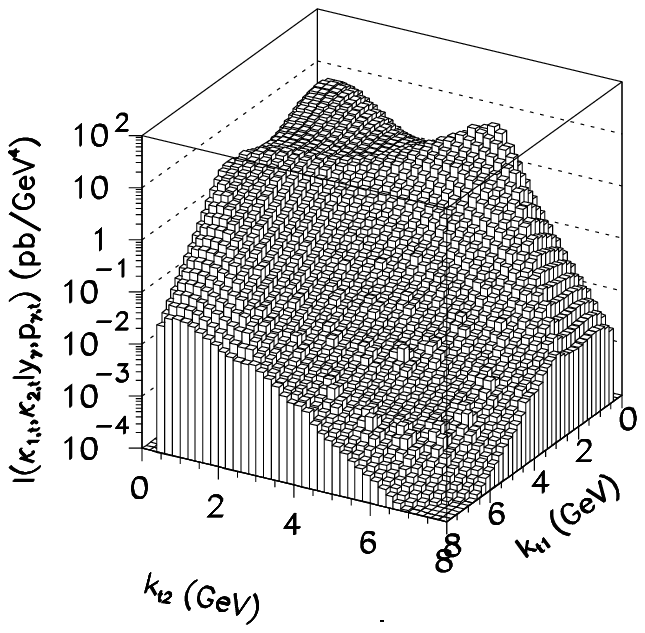}
\caption[*]{\it
Integrand under $k_{1,t}$ and $k_{2,t}$
in the invariant cross section formula for
proton-proton scattering at
$\sqrt{s} = 63$ GeV and y= 0, $p_{t,\gamma}$ = 5 GeV
and Kwieci\'nski UGDF ($b_0$ = 1 GeV$^{-1}$, $\mu^2$ = 100 GeV$^2$).
Off-shell matrix elements for gluons are used.
(a) $q \bar q$, (b) $\bar qq$, (c) $gq$, (d) $qg$.
\label{fig:kwiec_kt1_kt2_w63}
}
\end{center}
\end{figure}

In Fig.\ref{fig:kwiec_kt1_kt2_w63} we show a similar map for the
Kwieci\'nski distributions. In this case the factorization scale is
fixed for $\mu^2$ = 100 GeV$^2$. 

All kinematical variables are exactly the same as in 
the previous cases. The integrand is rather
similar to the one for the KMR UPDFs, except that the first maximum
at $k_{1,t}, k_{2,t} \approx$ 0 is somewhat broader.
We shall see consequences of the different integrands when discussing 
transverse momentum dependence of the photon inclusive cross sections.


\subsection{Off-shell effects}

Let us quantify the kinematical off-shell effect by defining
the following quantities:
\begin{eqnarray}
\begin{split}
&R_{qg}(k_{1,t}^2,k_{2,t}^2) &=&
\frac{I_{qg}^{off-shell}(k_{1,t}^2,k_{2,t}^2)}
     {I_{qg}^{on-shell}(k_{1,t}^2,k_{2,t}^2)}
\; , \\
&R_{gq}(k_{1,t}^2,k_{2,t}^2) &=&
\frac{I_{gq}^{off-shell}(k_{1,t}^2,k_{2,t}^2)}
     {I_{gq}^{on-shell}(k_{1,t}^2,k_{2,t}^2)}
\; .
\label{ratio_off-shell-to-on-shell}
\end{split}
\end{eqnarray}

\begin{figure}[!htb] 
\begin{center}
\includegraphics[width=.23\textwidth]{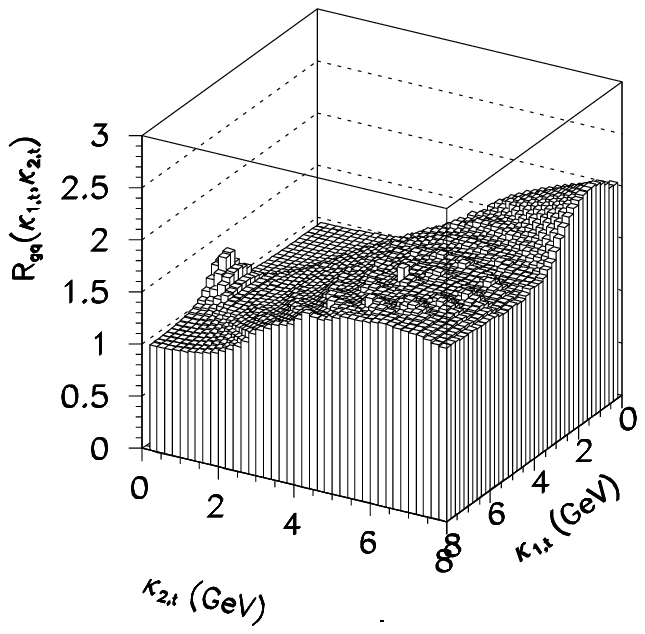}
\includegraphics[width=.23\textwidth]{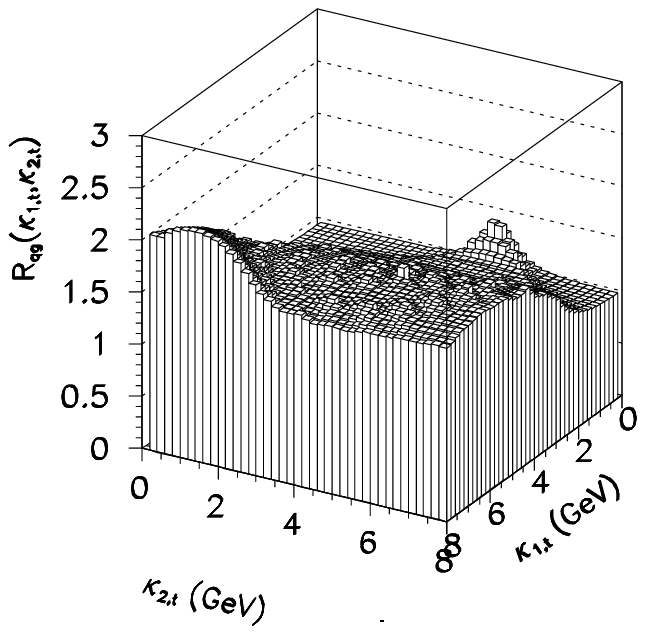}
\includegraphics[width=.23\textwidth]{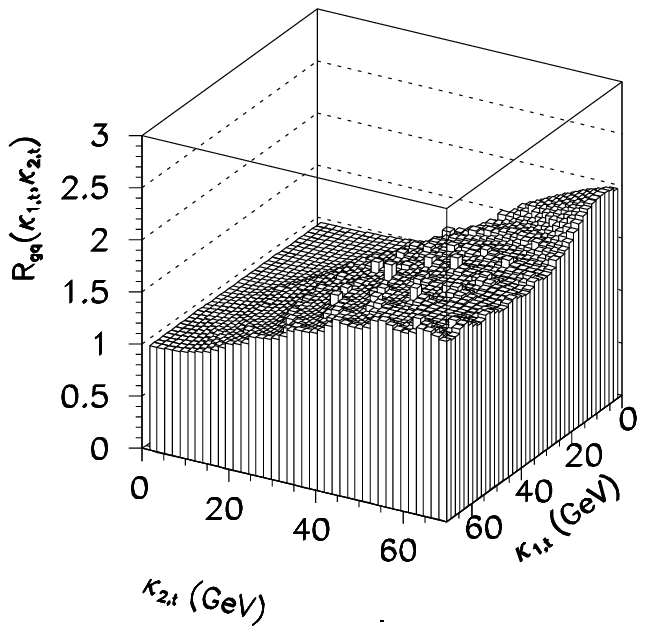}
\includegraphics[width=.23\textwidth]{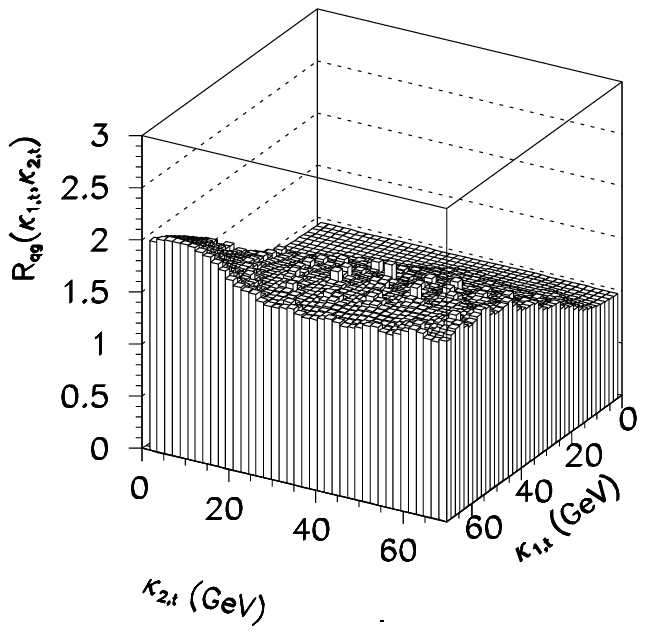}
\caption[*]{
Off-shell effects for the dominant mechanisms for
$pp$ scattering at $\sqrt{s}$ = 63 GeV, y = 0, $p_t$ = 5 GeV (upper panels) and
$p\bar p$ scattering at $\sqrt{s}$ = 630 GeV, y = 0, $p_t$ = 50 GeV
(lower panels).
\label{fig:off-shell-to-on-shell}
}
\end{center}
\end{figure}

In Fig.\ref{fig:off-shell-to-on-shell} we present results for
$R_{gq}$ (left panels) and $R_{qg}$ (right panels)
for proton-proton scattering at W = 63 GeV
($y_{\gamma}$ = 0, $p_{\gamma,t}$ = 5 GeV) and
proton-antiproton scattering at W = 630 GeV
($y_{\gamma}$ = 0, $p_{\gamma,t}$ = 50 GeV).
When $k_{1,t}, k_{2,t} \to$ 0 the off-shell effects dissapear,
i.e. the ratio becomes unity. The larger transverse momenta of gluons
the larger the off-shell effect is. Therefore one may expect 
a related enhancement
of the photon inclusive cross section when the UGDFs with large
transverse momentum spread are used.

\begin{figure}[!htb] 
\begin{center}
\includegraphics[width=.4\textwidth]{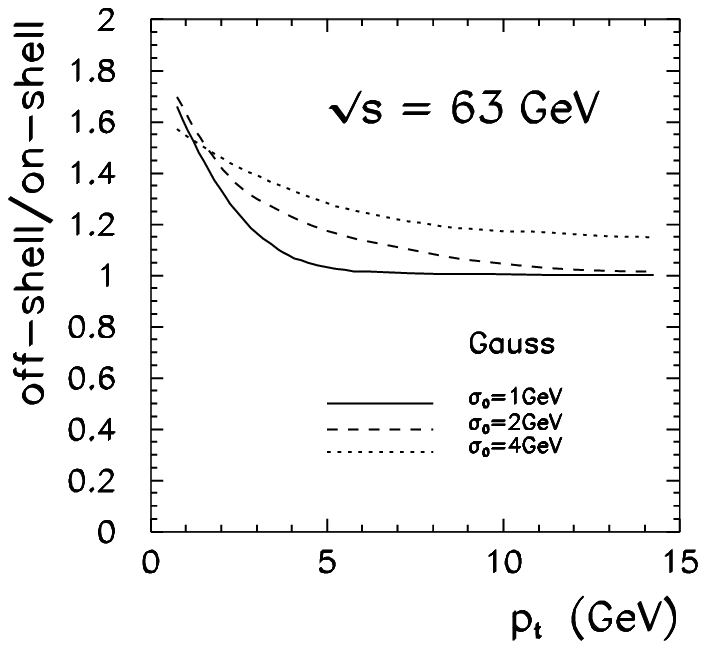}
\includegraphics[width=.4\textwidth]{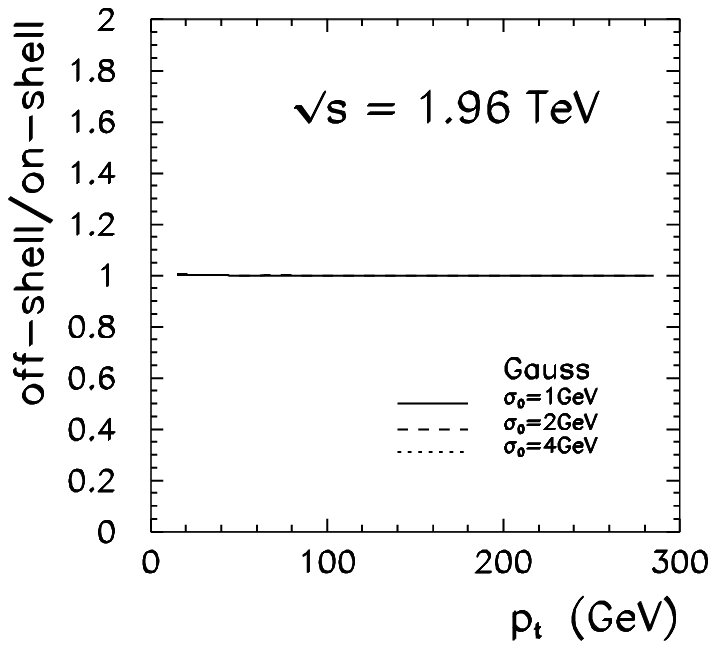}
\caption[*]{
The ratio of inclusive cross sections with off-shell and on-shell
matrix elements with Gaussian UPDFs and different values of
the parameter $\sigma_0$ as a function of photon transverse momentum.
\label{fig:off-shell-to-on-shell_pt_gauss}
}
\end{center}
\end{figure}

In Fig.\ref{fig:off-shell-to-on-shell_pt_gauss} we show 
the ratio of the
inclusive cross sections obtained with off-shell and on-shell matrix
elements as a function of photon transverse momentum.
In this calculation the Gaussian distributions were used with different
values of the $\sigma_0$ parameter.  The bigger $\sigma_0$ the larger
the enhancement due to the off-shell effects.

\begin{figure}[!htb] 
\begin{center}
\includegraphics[width=.4\textwidth]{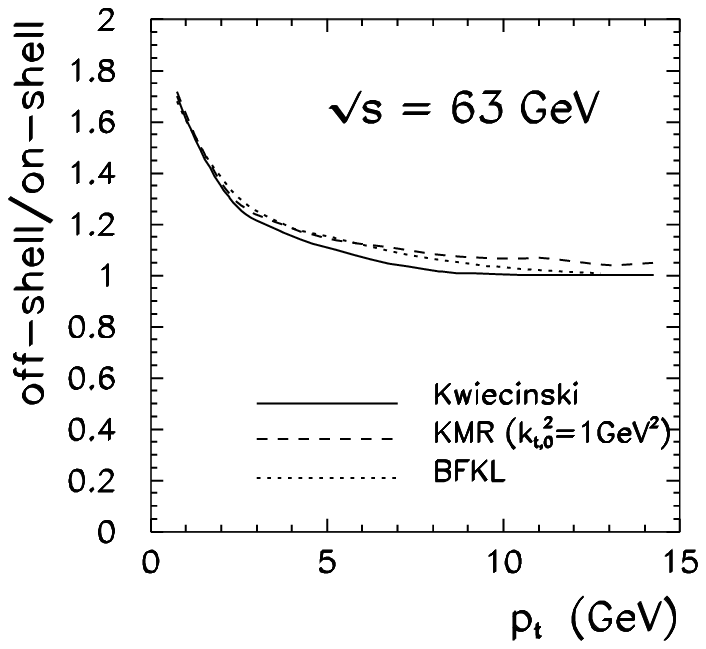}
\includegraphics[width=.4\textwidth]{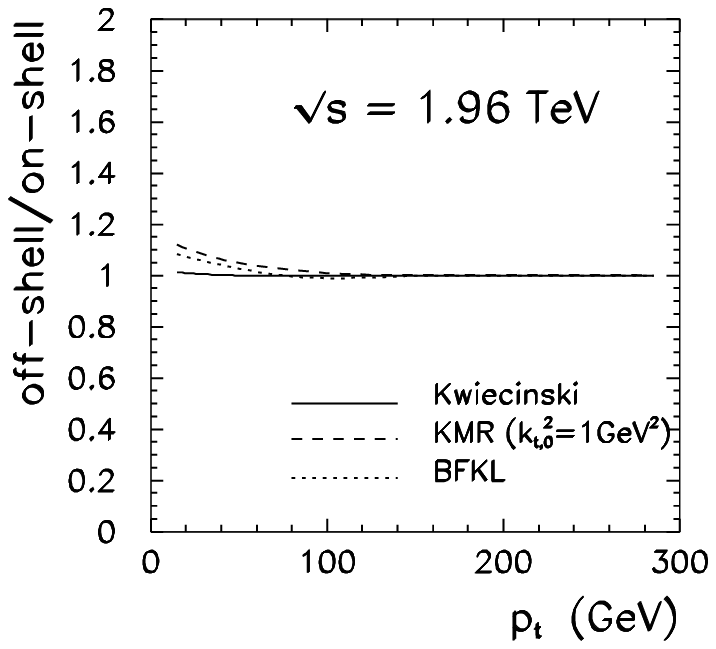}
\caption[*]{
The ratio of inclusive cross sections with off-shell and on-shell
matrix elements for different UPDFs as a function of photon transverse
momentum.
\label{fig:off-shell-to-on-shell_pt_UPDFs}
}
\end{center}
\end{figure}

In Fig.\ref{fig:off-shell-to-on-shell_pt_UPDFs}
we show similar enhancement for a few representative UPDFs discussed
in section 2. 
The biggest enhancement is obtained with the KMR and BFKL
distributions, i.e. those which have the biggest gluon 
transverse momentum spread.
In general, the bigger photon transverse momentum, the smaller
the enhancement. We conclude that at larger photon transverse momenta one
can use standard on-shell matrix elements.

\subsection{Photon transverse momentum distributions}

\begin{figure}[!htb] 
\begin{center}
\includegraphics[width=.4\textwidth]{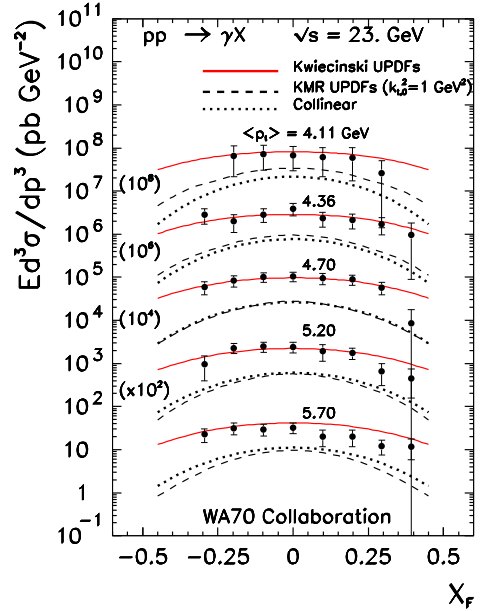}
\caption[*]{
Invariant cross section for direct photons for $\sqrt s$ = 23 GeV
as a function of Feynman $x_F$ for different 
bins of transverse momenta.
In this calculation of shell matrix elements for subprocesses with
gluons were used. The Kwieci\'nski UPDFs were calculated with
the factorization scale $\mu^2$ = 100 GeV$^2$. The theoretical results
are compared with the WA70 collaboration data \cite{data_WA70}.
\label{fig:inv_cs_w23}
}
\end{center}
\end{figure}

\begin{figure}[!htb] 
\begin{center}
\includegraphics[width=.36\textwidth]{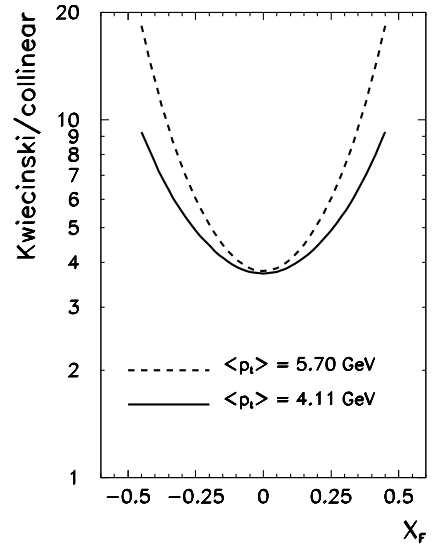}
\caption[*]{
The ratio of the invariant cross sections as a function of $x_F$
for $\sqrt s$ = 23 GeV.
The Kwieci\'nski UPDFs were calculated with
the factorization scale $\mu^2$ = 100 GeV$^2$.
\label{fig:kwiec_to_coll_xf}
}
\end{center}
\end{figure}

Let us start the analysis from the lowest energies.
In Fig.\ref{fig:inv_cs_w23} we show inclusive invariant cross section
as a function of Feynman $x_F$ for several experimental values of
photon transverse momenta as measured by the WA70 collaboration.
It is well known that the collinear approach (dotted line) fails to
describe the low transverse momentum data by a sizeable factor of 4 or
even more. Also the $k_t$-factorization result with the KMR UPDFs
(dashed line) underestimate the low-energy data. In contrast,
the Kwieci\'nski UPDFs (solid line) describe the WA70 collaboration
data almost perfect. In order to illustrate the situation in
Fig.\ref{fig:kwiec_to_coll_xf} we show the ratio
\begin{equation}
\frac{d \sigma^{Kwiec}}{dy d^2p_t}(x_F,p_t) \bigg/
\frac{d \sigma^{coll}}{dy d^2p_t}(x_F,p_t)
\label{kwiec_to_coll}
\end{equation}
as a function of Feynman $x_F$ for $p_t$ = 4.11 GeV and 
$p_t$ = 5.70 GeV.
This figure shows that the enhancement factor 
strongly depends on $x_F$.
Such an enhancement is required by the experimental data 
as can be seen by inspection of the previous figure.

\begin{figure}[!htb] 
\begin{center}
\includegraphics[width=.4\textwidth]{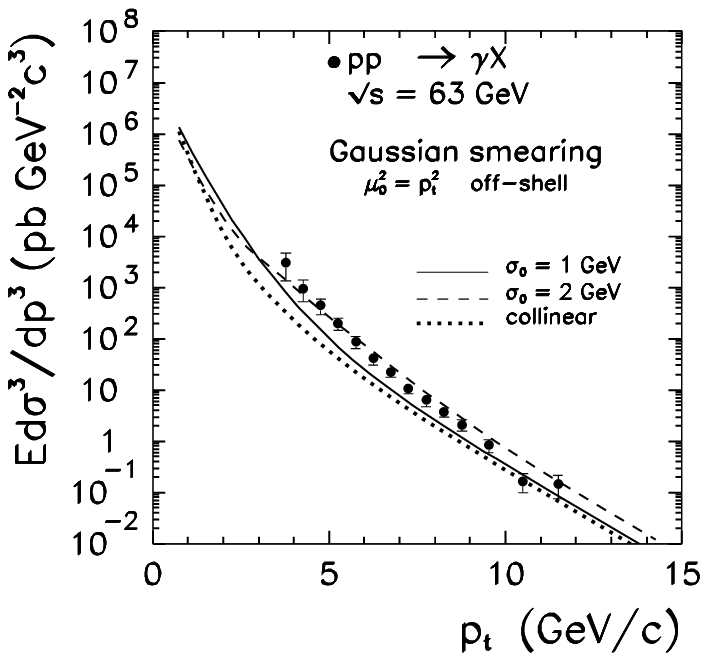}
\includegraphics[width=.4\textwidth]{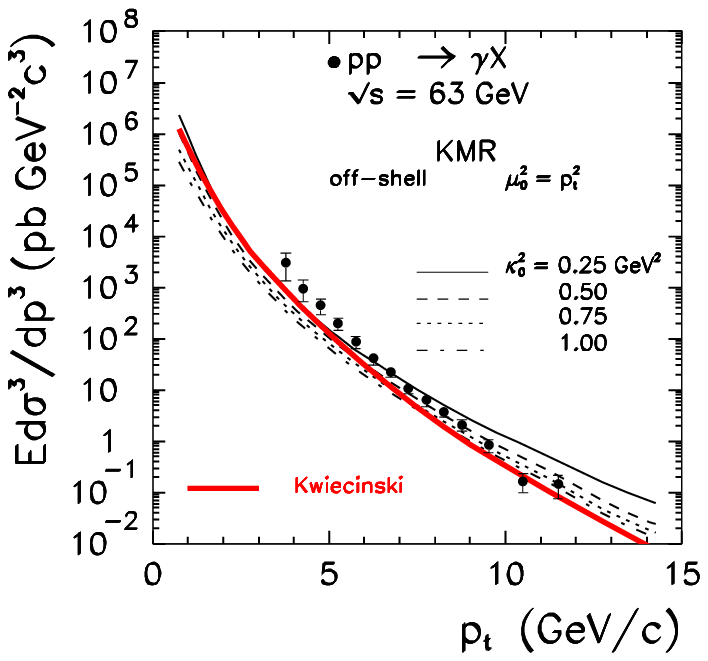}
\caption[*]{
Invariant cross section for direct photons for
$\sqrt s$ = 63 GeV and $y$ = 0 as a function of photon transverse momentum.
In this calculation off-shell matrix element for gluons were used.
The experimental data of the R806 collaboration are
taken from Ref.\cite{data_R806}.
(a) Gaussian smearing ($\sigma_0$ = 1, 2 GeV) versus collinear approach
(b) standard KMR prescription,
\label{fig:inv_cs_w63}
}
\end{center}
\end{figure}

In Fig.\ref{fig:inv_cs_w63} we show invariant cross section
for direct photon production as a function of photon transverse
momentum for photon rapidity $y$ = 0 and $\sqrt{s}$ = 63 GeV.
The results obtained with the KMR UPDFs strongly depend on the 
value of the parameter $k_{t,0}^2$. 
The larger the parameter $k_{t,0}^2$, the smaller
cross section. This means that even at large photon transverse
momenta the nonperturbative effects (small $k_t$'s) play 
an important role. This can be better understood via inspection
of the two-dimensional maps $k_{1,t} \times k_{2,t}$
shown in Fig.\ref{fig:kmr_kt1_kt2_w63} and is related to the second
and third local maxima which give a significant contribution
to the invariant cross section. 

In principle, one could try to find
the parameter $k_{t,0}^2$ by confronting the theoretical 
results with experimental data. If the parameter is adjusted 
to larger transverse momenta there is a deficit at smaller 
transverse momenta compared to the ISR data \cite{data_R806}.

\begin{figure}[!htb] 
\begin{center}
\includegraphics[width=.4\textwidth]{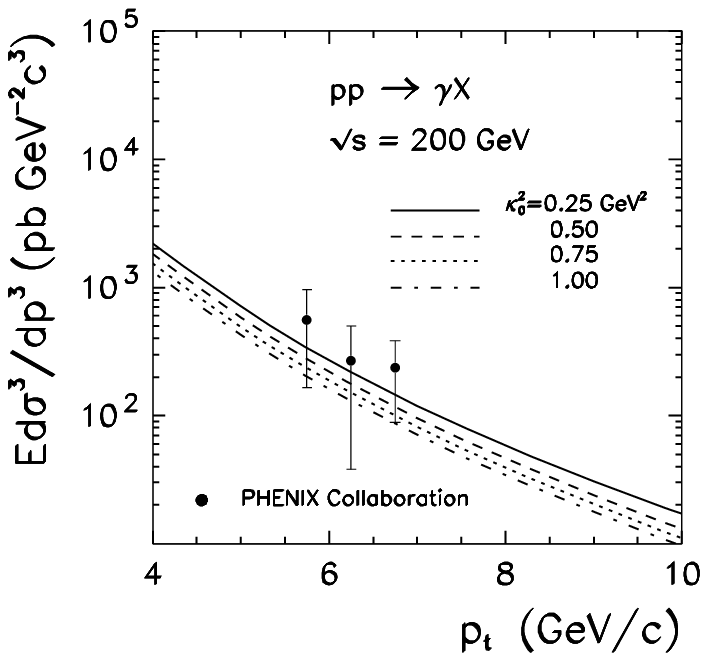}
\includegraphics[width=.4\textwidth]{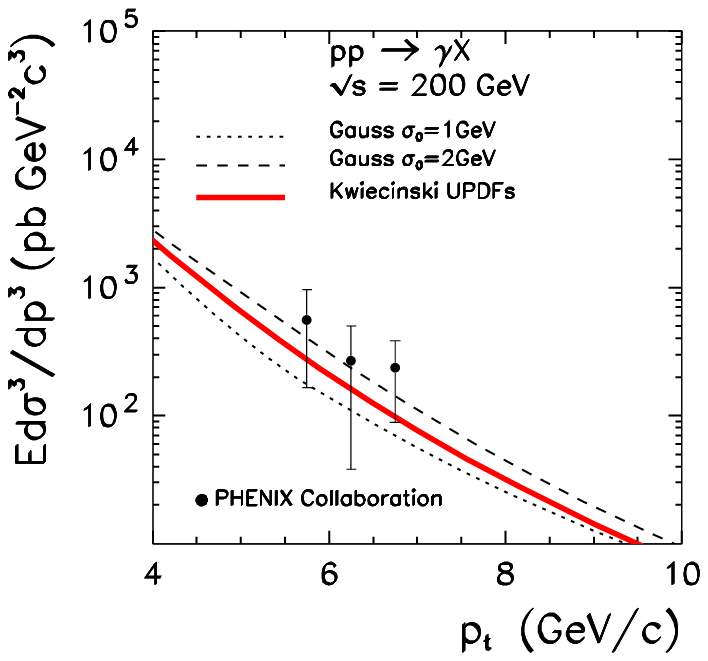}
\caption[*]{
Invariant cross section for direct photons for
$\sqrt s = 200$ GeV and $y$ = 0.
In this calculation off-shell matrix element for gluons were used.
The experimental data of the PHENIX collaboration
are taken from Ref.\cite{data_PHENIX}.
(a) standard KMR prescription
(b) Gaussian smearing ($\sigma_0$ = 1, 2 GeV) versus Kwieci\'nski
    and collinear approach
\label{fig:inv_cs_rhic}
}
\end{center}
\end{figure}

In Fig.\ref{fig:inv_cs_rhic} we compare our results with recent
proton-proton RHIC data \cite{data_PHENIX}. 
Here only low transverse-momenta of photons were measured.
The results obtained with the Gaussian UPDFs
strongly depend on the value of the $\sigma_0$ parameter which is
not surprising for the low transverse momenta.
The Kwieci\'nski distributions give fairly good description 
of the PHENIX data.
 
\begin{figure}[!htb] 
\begin{center}
\includegraphics[width=.39\textwidth]{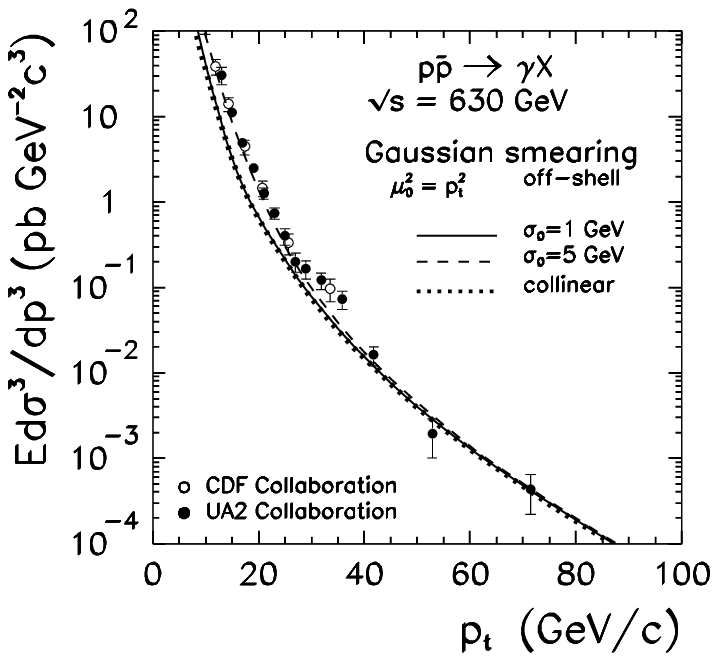}
\includegraphics[width=.39\textwidth]{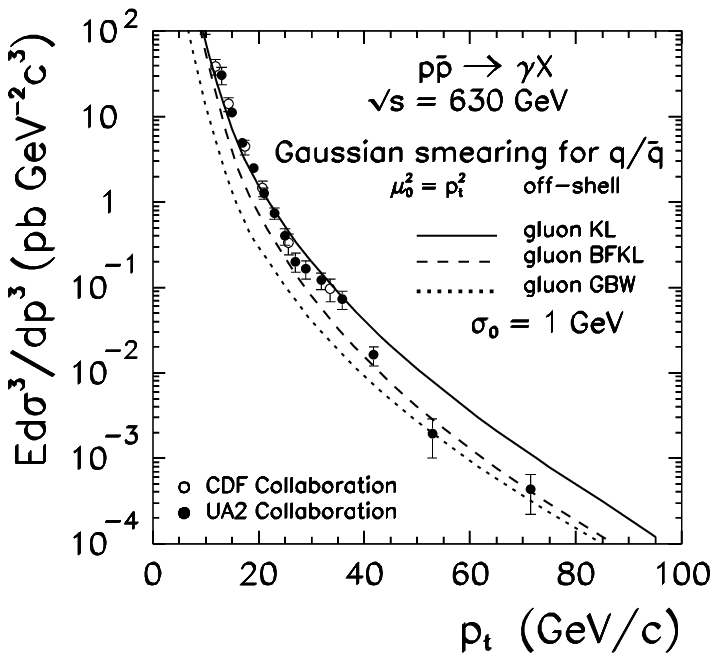}
\includegraphics[width=.39\textwidth]{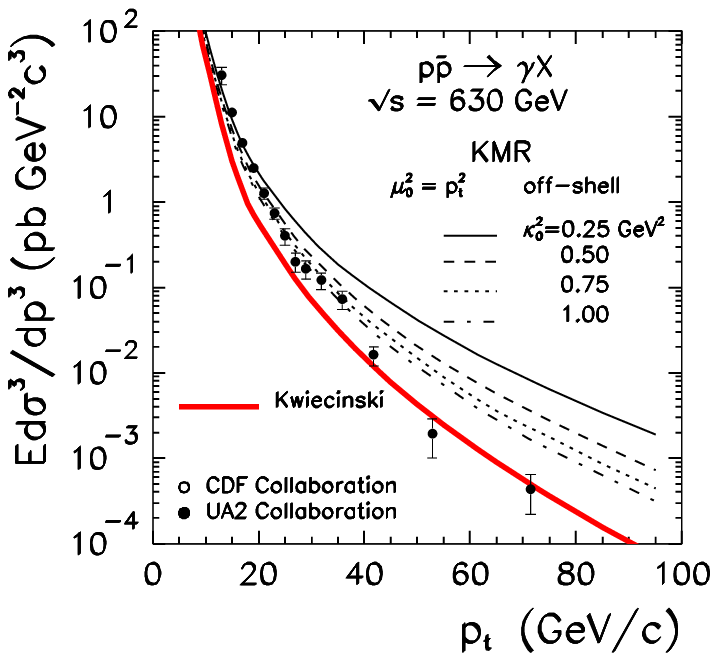}
\caption[*]{
Invariant cross section for direct photons for
$\sqrt s = 630$ GeV. In this calculation off-shell matrix element for
gluons were used. The experimental data of the UA2 collaboration
were taken from Ref.\cite{data_UA2}.
(a) Gaussian smearing, 
(b) quark/antiquarks: Gaussian smearing ($\sigma_0$ = 1 GeV), 
gluons: KL, BFKL and GBW,
(c) standard KMR prescription.
\label{fig:inv_cs_w630}
}
\end{center}
\end{figure}

\begin{figure}[!htb] 
\begin{center}
\includegraphics[width=.4\textwidth]{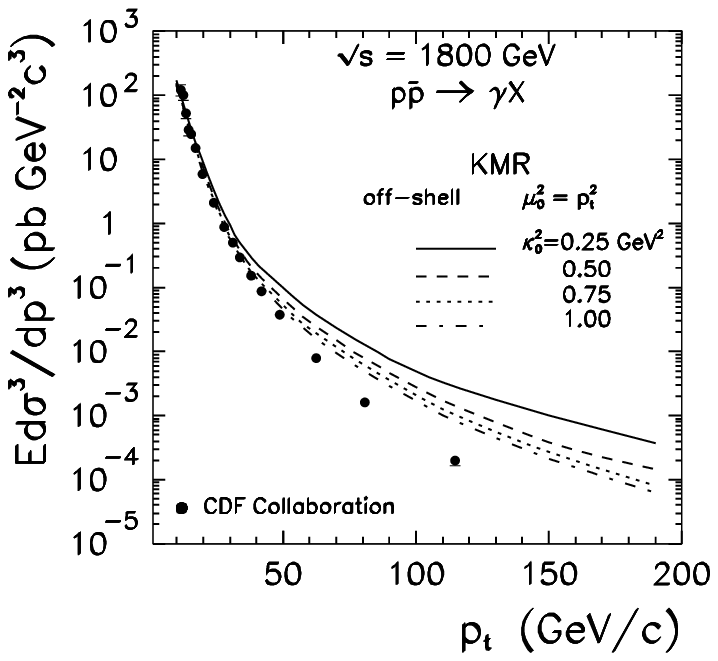}
\includegraphics[width=.4\textwidth]{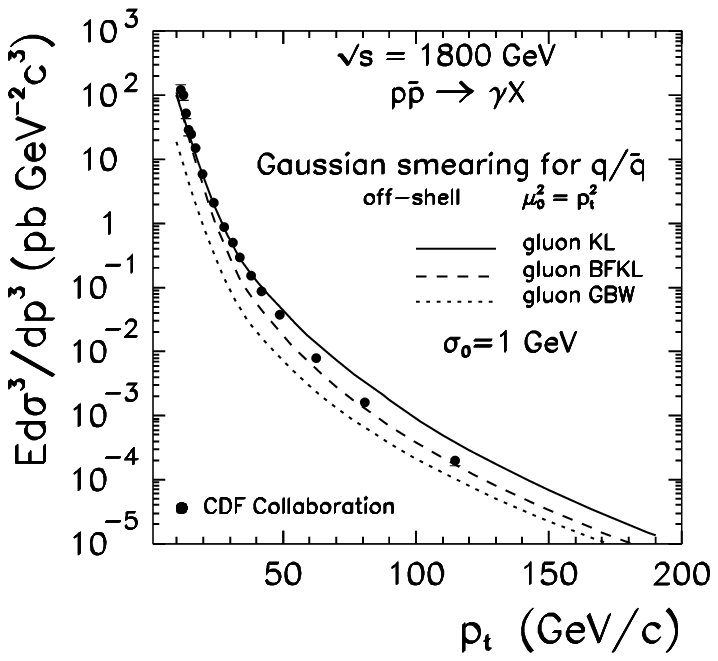}
\caption[*]{
Invariant cross section for direct photons for
$\sqrt s = 1800$ GeV. In this calculation off-shell matrix element for
gluons was used.
(a) standard KMR prescription,
(b) quarks/antiquarks: Gaussian smearing ($\sigma_0$ = 1 GeV), 
gluons: KL, BFKL and GBW.
\label{fig:inv_cs_w1800}
}
\end{center}
\end{figure}

\begin{figure}[!htb] 
\begin{center}
\includegraphics[width=.4\textwidth]{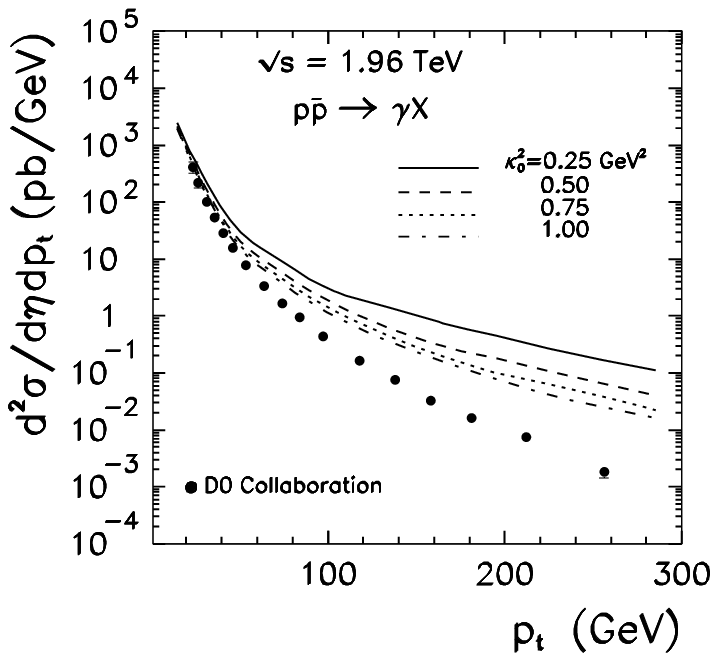}
\includegraphics[width=.4\textwidth]{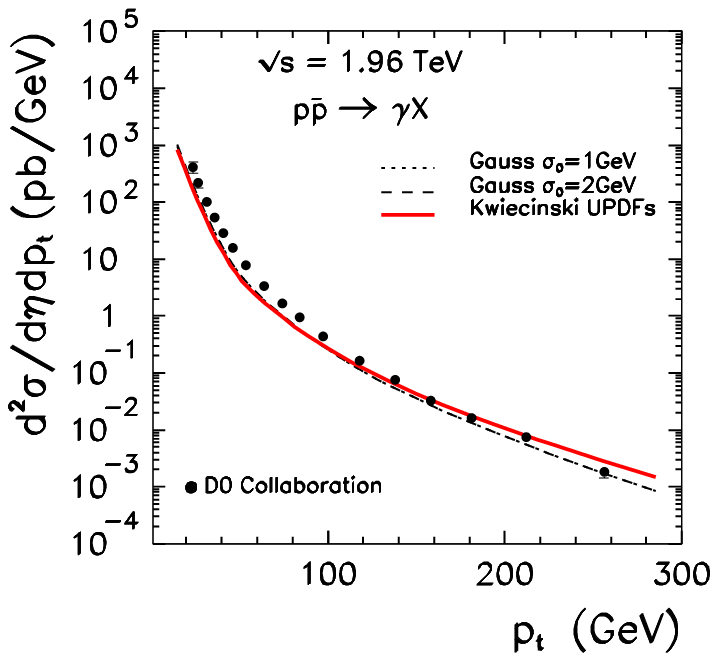}
\caption[*]{
Cross section for direct photons for
$\sqrt s = 1.96$ TeV. In this calculation off-shell matrix element for
gluons were used.
(a) standard KMR prescription,
(b) Gaussian smearing ($\sigma_0$ = 1,2 GeV) versus Kwieci\'nski UPDFs. 
\label{fig:inv_cs_w1960}
}
\end{center}
\end{figure}

In contrast to the ``low energy'' data, there is no deficit for the KMR
UPDFs at larger energies as can be seen from
Figs. \ref{fig:inv_cs_w630}, \ref{fig:inv_cs_w1800} and
\ref{fig:inv_cs_w1960}. 
The KMR UPDFs, however, strongly overestimate
the experimental data at large photon transverse momenta. This is especially
visible for proton-antiproton collisions at W = 1.96 TeV when compared
with recent Tevatron (run 2) data \cite{data_D0_w1960}.
Figs.\ref{fig:inv_cs_w1800} and \ref{fig:inv_cs_w1960} show that
the unintegrated parton distribution approach with the KMR UPDFs is
clearly inconsistent with the standard collinear approach at large
transverse momenta. This is caused by the presence of large-$k_t$ tails
(of the 1/$k_t$ type) in the KMR UPDFs.
It is not the case for the Gaussian and Kwieci\'nski UPDFs which 
seem to converge to the standard collinear result at large photon
transverse momenta. In this respect the latter UPDFs seems preferable.

\subsection{Direct photons at LHC}
 
Up to now we have confronted our results with the existing experimental
data from SPS, ISR, RHIC, S$p \bar p$S and Tevatron.
In a not too distant future one may expect experimental data from
a newly constructed LHC.

\begin{figure}[!htb] 
\begin{center}
\includegraphics[width=.4\textwidth]{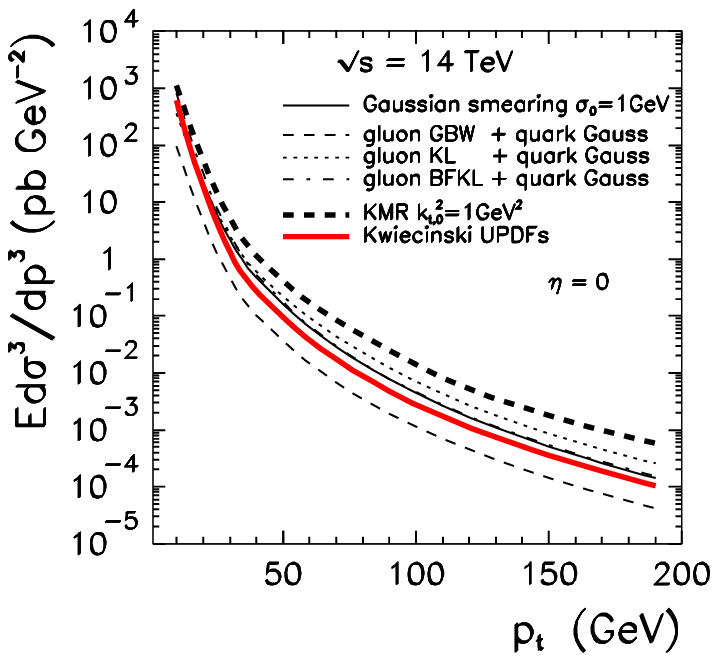}
\includegraphics[width=.4\textwidth]{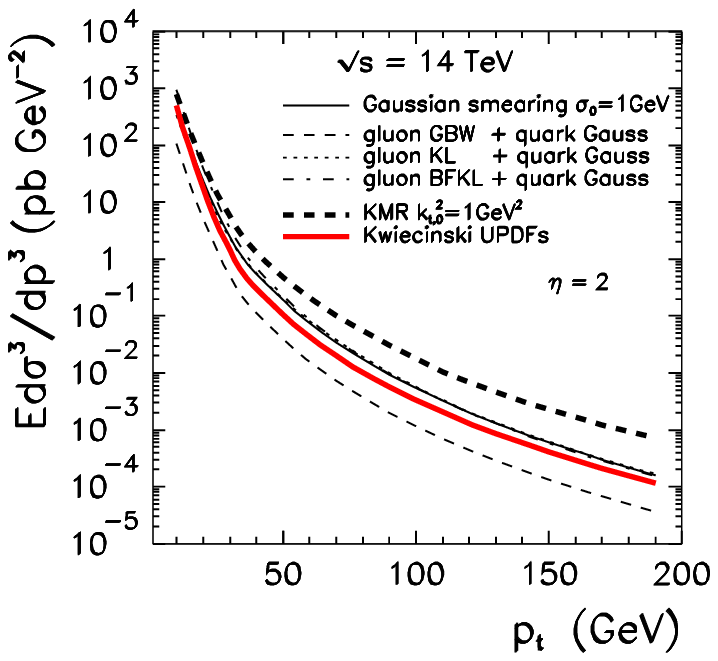}
\includegraphics[width=.4\textwidth]{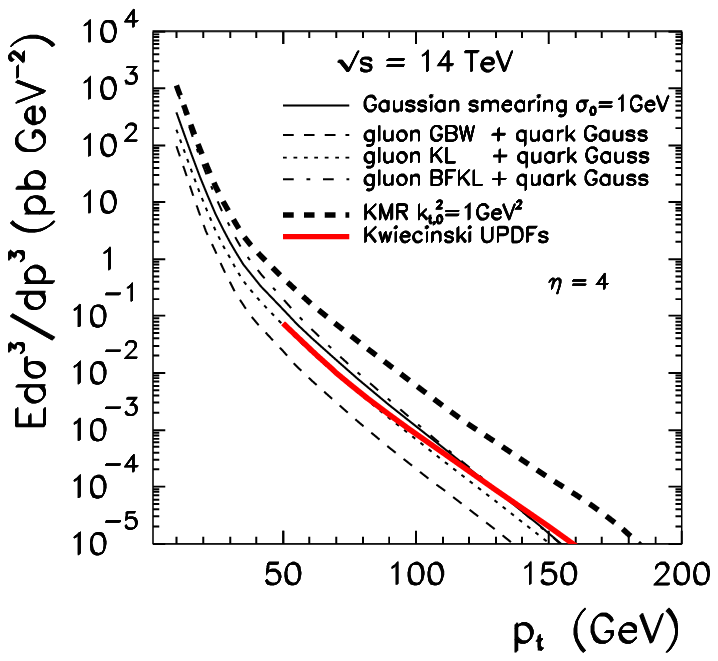}
\caption[*]{
Predictions for LHC ($\sqrt{s}$ = 14 TeV) for different
photon rapidities y = 0, 2, 4. We compare results
for different UPDFs: KMR($k_{t,0}^2$ = 1 GeV$^2$), GRV+Gaussian
smearing ($\sigma_0$ = 1 GeV) and BFKL, KL and GBW UGDFs with
quarks/antiquarks smeared by a Gaussian distribution
($\sigma_0$ = 1 GeV).
\label{fig:LHC_predictions_pt_fixy}
}
\end{center}
\end{figure}

In Fig.\ref{fig:LHC_predictions_pt_fixy} we present transverse momentum
dependence of the invariant cross section for
the proton-proton collisions at W = 14 TeV for different values
of photon rapidities. 
Different UPDFs lead to quite different results
at the LHC energies.
Therefore future measurements at LHC should give a chance to verify
different UPDFs discussed here as well as others to be constructed in
the future.

\begin{figure}[!htb] 
\begin{center}
    \includegraphics[width=.35\textwidth]{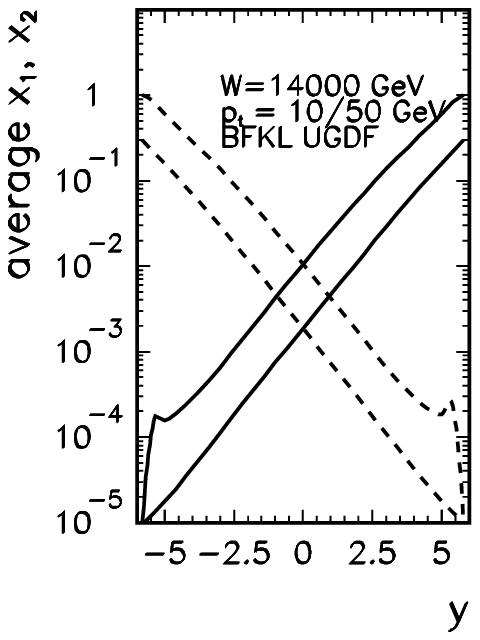}
\caption[*]{
Average values of $x_1$ (solid) and $x_2$ (dashed) as a 
function of photon rapidity.
The borders of the bands shown correspond to $p_t$ = 10 GeV 
(lower limit)
ant $p_t$ = 50 GeV (upper limit).
\label{fig:ave_x1x2_bfkl}
}
\end{center}
\end{figure}

In Fig.\ref{fig:ave_x1x2_bfkl} we display average values of $x_1$ and
$x_2$ of partons participating in a hard subprocess for two
values of the photon transverse momentum $p_t$ = 10, 50 GeV.
\begin{figure}[htb] 
\begin{center}
\includegraphics[width=.23\textwidth]{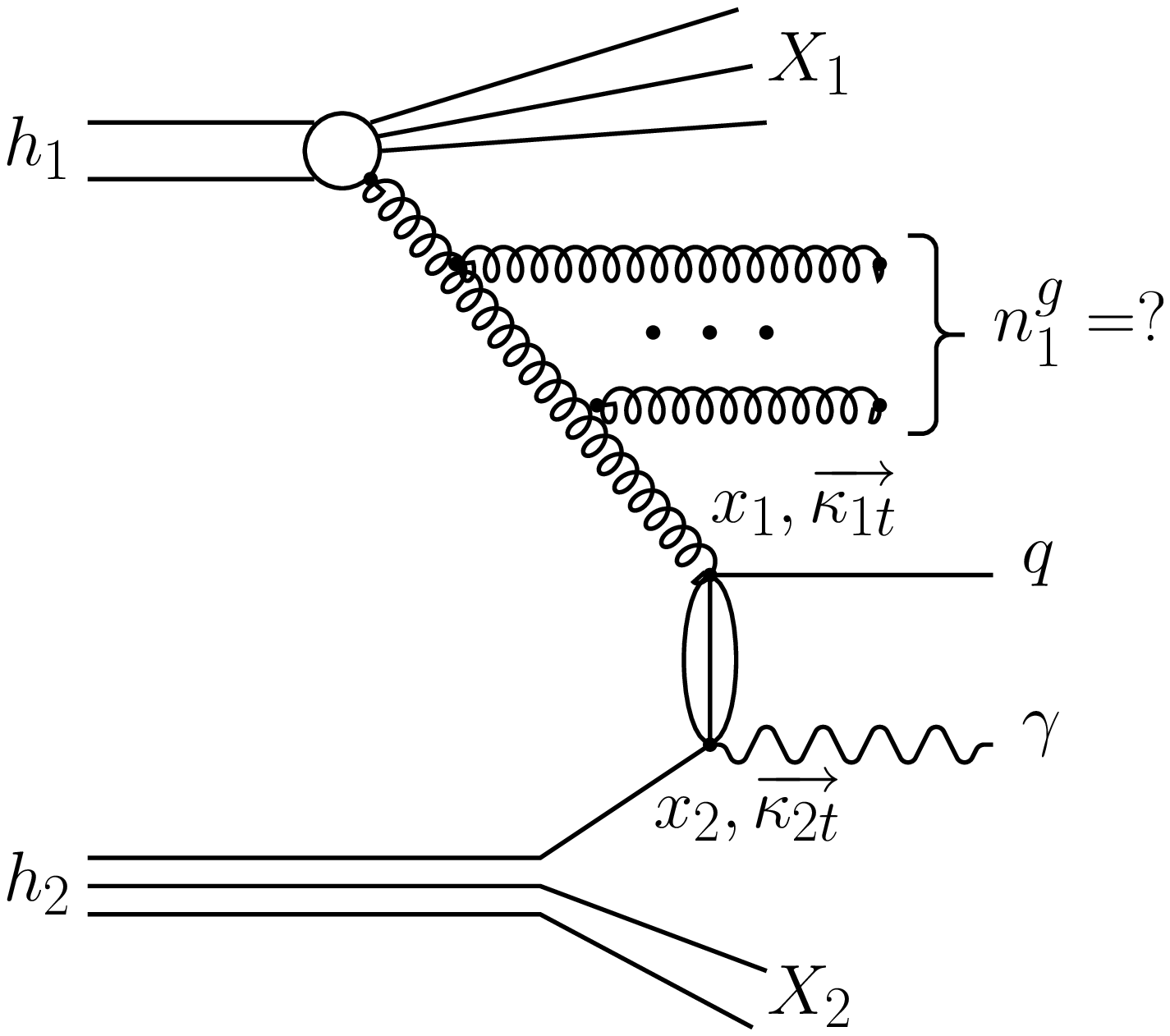}
\includegraphics[width=.23\textwidth]{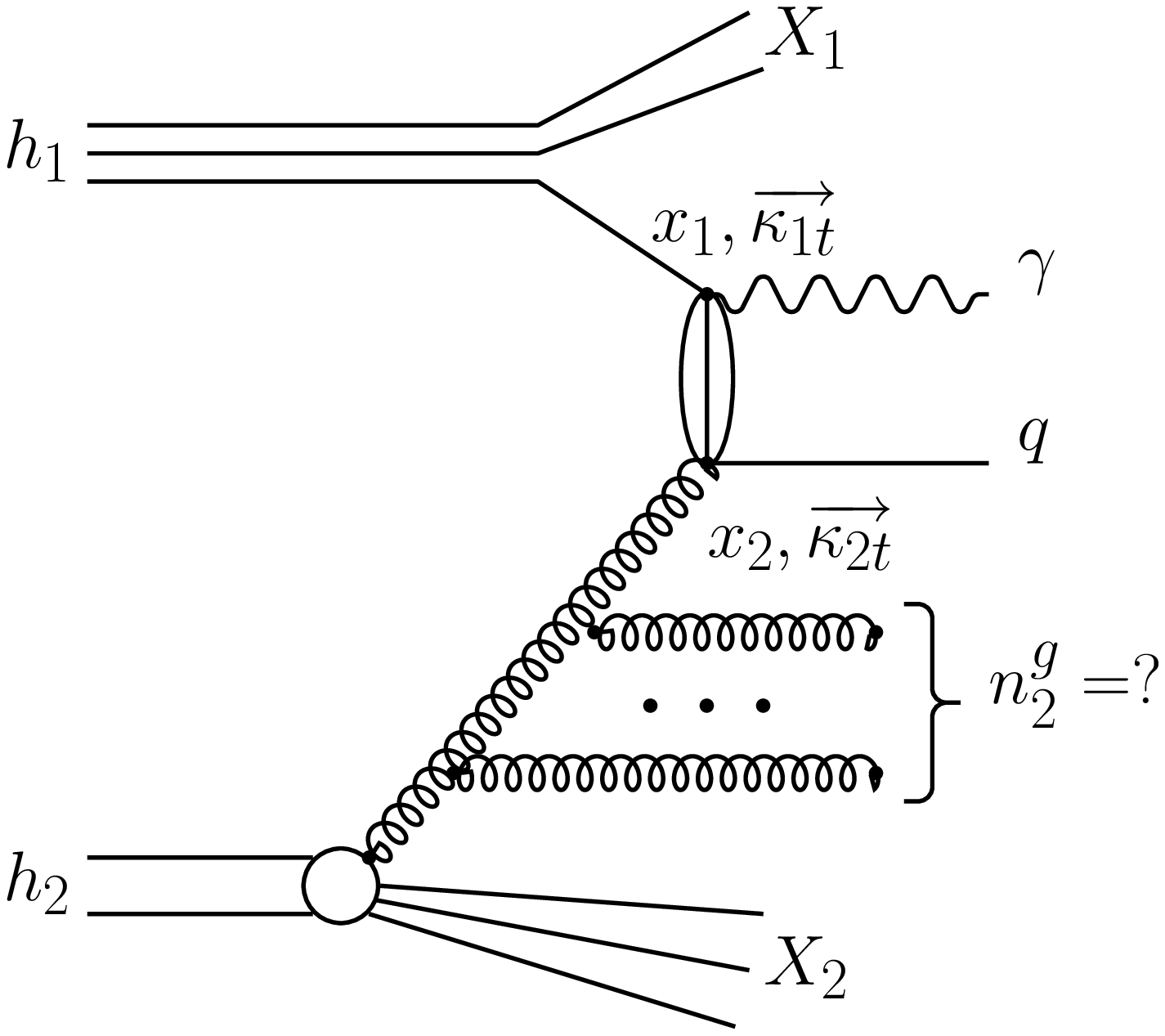}
\caption[*]{
The dominant mechansisms of photon production
at forward (left panel) and backward (right panel) rapidities.
\label{fig:ladder_diagrams}
}
\end{center}
\end{figure}
At large rapidities either $x_1 \gg x_2$ or $x_1 \ll x_2$.
Then one expects the dominance of $q(valence)g$ or $gq(valence)$ hard
processes (see Fig.\ref{fig:ladder_diagrams}). 
At such small values of $x$
the evolution effects for UGDFs are expected to be very important.
In addition, one expects rather small transverse momenta of large-x
valence quarks, much smaller than transverse momenta of
the associated small-x gluons.

\begin{widetext}
\begin{figure}[htb] 
\begin{center}
    \includegraphics[width=.85\textwidth]{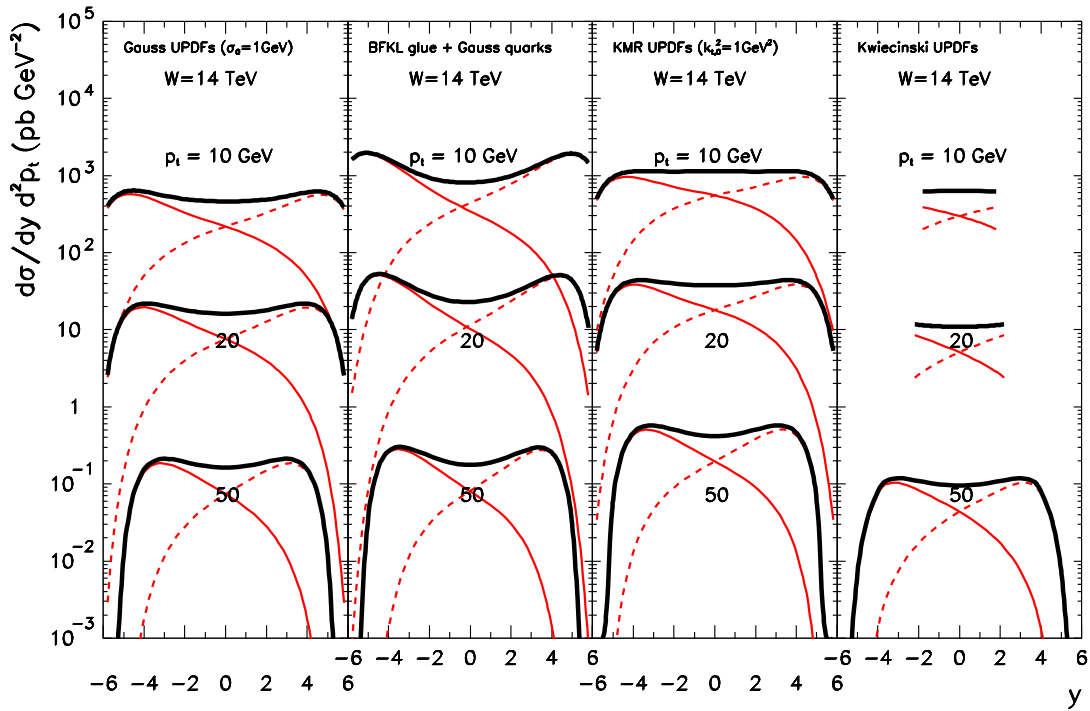}
\caption[*]{
The rapidity dependence of the invariant cross section for a few
values of photon transverse momenta ($p_t$ = 10, 20, 50 GeV).
The contribution of $g q$ (solid) and $q g$ (dashed) subprocesses
are shown separately for each value of the transverse momentum.
\label{fig:LHC_predictions_y_fixpt}
}
\end{center}
\end{figure}
\end{widetext}

In Fig.\ref{fig:LHC_predictions_y_fixpt} we show the dependence
of the invariant cross section on photon rapidity for fixed values of
the photon transverse momenta. 
\clearpage
Results obtained with different
UGDFs differ significantly in the region of large rapidities.

In the case of the Kwieci\'nski UPDFs at small transverse momenta
($p_t=10,20$ GeV) only limited part of the full curve is shown.
This is dictated by a purely technical cut on the longitudinal 
momentum fraction $x > 10^{-4}$
when constructing interpolation maps of the Kwieci\'nski UPDFs.
As can be seen from Fig.\ref{fig:ave_x1x2_bfkl} large $|y|$ require
very small $x_1$ or $x_2$, sometimes smaller than $10^{-4}$.
Furthermore the Kwieci\'nski distributions are not expected to be
reliable at such small values of longitudinal momentum fractions.
Therefore the range of application of the Kwieci\'nski UPDFs at LHC
is limited in rapidity and photon transverse momentum.
The larger photon transverse momentum, the broder range of application
in photon rapidity.

In conclusion, the region of large rapidities
($|y| >$ 3), discussed in this section, seems an appropriate 
place to test models of UPDFs. 
It is not clear to us if any of the LHC detectors can register
the large-energy forward/backward photons. The CASTOR detector
associated with the CMS detector is a potential option.



\section{Conclusions}

The inclusive cross section for prompt photon production has been
calculated for different incident energies from SPS to LHC within
the formalism of unintegrated parton distributions.
Different models of UPDFs lead to rather different results.
The Kwieci\'nski distributions provide the best description of
experimental data in the broad range of incident energies.
The existing experimental data test UPDFs down to x = 10$^{-3}$, i.e.
in the region of intermediate longitudinal momentum fractions
adequate for application of the Kwieci\'nski equations.
Inclusion of the QCD evolution effects and especially
their effect on initial parton transverse momenta allowed to solve 
the long-standing problem of theoretical understanding 
the low energy and low transverse momentum data for direct 
photon production.

As a buy-product we have analyzed momentum sum rule for different UPDFs.
We have found that the KMR UPDFs violate naive number sum rules.
The same distributions lead to an interesting interplay of soft (small
gluon $k_t$'s) and hard (large gluon $k_t$'s) regions of UPDFs.
Even at large photon transverse momenta this interplay
causes a huge enhancement as compared to collinear approach,
quite inconsistent with the experimental data at large photon
transverse momenta.

We have presented predictions for LHC based on several UPDFs with
special emphasis on large rapidity region. Here different UPDFs lead
to quite different predictions. Therefore we conclude that this region
can be very useful to test different UPDFs.

\section{Acknowledgments}
We are indebted to Artem Lipatov for an interesting and instructive
discusion concerning their work on direct photon production.
This paper was partially supported by the grant of the Polish Ministry
of Scientific Research and Information Technology number 1 P03B 028 28.

\section{Appendix A}

The moving with center-of-mass hadron-hadron energy maxima
for the KMR distributions cause that the integration in
$d k_{1,t}$ and $d k_{2,t}$ is not very efficient,
especially for large $p_t$.
In order to make the integration more efficient we perform
a change of the variables in the integration
$d k_{1,t} d k_{2,t} \to dy_1 dy_2$,
where $y_i = \mathrm {log}_{10}(k_{i,t}^2)$. Then
\begin{eqnarray*}
y_i &=& \mathrm {log}_{10}(k_{i,t}^2) \;\;
\to \;\; k_{i,t}^2 = 10^{y_i}\; ,  \\
dy_i &=& \frac{1}{k_{i,t}^2} \frac{1}{\mathrm {ln}(10)}
2 k_{i,t} d k_{i,t} \; ,  \\
k_{i,t} d k_{i,t} &=& \frac{1}{2}10^{y_i}\mathrm {ln}(10)dy_i  
\label{transfo}
\end{eqnarray*}
which gives
\begin{eqnarray*}
\int k_{1,t} d k_{1,t} k_{2,t} d k_{2,t} &=&
\frac{1}{4} \int 10^{y_1}\mathrm {ln}(10) dy_1 
                     10^{y_2}\mathrm {ln}(10) dy_2  \\
&=& \frac{1}{4} \int 10^{(y_1+y_2)}\mathrm {ln}^2 (10)dy_1 dy_2 \; . 
\label{new_differential_element}
\end{eqnarray*}
Then the invariant cross section for the production of the photon 
(associated with a parton) can be writen as
\begin{widetext}
\begin{eqnarray*}
\sigma_{inv}(\eta_1,p_{1,t}) &\equiv& 
\frac{d\sigma(h_1h_2\to \gamma X)}{d\eta_1d^2p_{1,t}}\\ 
&=& \frac{1}{4}
\int d\phi_1 d\phi_2 \cdot 
10^{\mathrm {log}_{10}(k_{1,t}^2)}\mathrm {ln}(10)d\mathrm {log}_{10}(k_{1,t}^2)\cdot
10^{\mathrm {log}_{10}(k_{2,t}^2)}\mathrm {ln}(10)d\mathrm {log}_{10}(k_{2,t}^2)\cdot
d\eta_2 \\
&\cdot& 
\frac{1}{16\pi^2}\frac{1}{(x_1x_2s)^2}
\sum_{i,j,k} \overline{|\mathcal{M}_{ij \to \gamma k}|^2}
\frac{f_i(x_1,k_{1,t}^2)}{\pi}
\frac{f_j(x_2,k_{2,t}^2)}{\pi} \; .
\end{eqnarray*}
\end{widetext}
In the formula above $f_i(x_1,k_{1,t}^2)$ and $f_j(x_2,k_{2,t}^2)$
are unintegrated parton distribution functions. 
The invariant cross section can be formally written as
\begin{widetext}
\begin{equation*}
\sigma_{inv}(\eta_1,p_{1,t}) =
\int d \mathrm{log}_{10}(k_{1,t}^2) d \mathrm{log}_{10}(k_{2,t}^2) \;
I_{log-log}(\mathrm{log}_{10}(k_{1,t}^2), \mathrm{log}_{10}(k_{2,t}^2))
\label{integrand_loglog} \; .
\end{equation*}
\end{widetext}

\section{Appendix B}

The on-shell as well as off-shell matrix elements 
$\overline{|\mathcal{M}_{ij \to \gamma X}|^2}$
are taken into account for the following subprocesses
\begin{eqnarray*}
q \bar q &\to& \gamma g \;\; (on-shell) \; ,\\
\bar q q &\to& \gamma g \;\; (on-shell) \; ,\\
g      q &\to& \gamma q \;\; (on-shell, off-shell) \; ,\\
q      g &\to& \gamma q \;\; (on-shell, off-shell) \; .
\label{processes_included}
\end{eqnarray*}
When neglecting parton masses the on-shell matrix elements squared can be
written as \cite{Owens}
\begin{widetext}
\begin{eqnarray*}
\overline{|\mathcal{M}_{q \bar q \to \gamma X}|^2} 
&=&
\pi \alpha_{em}\sqrt{\alpha_{1,s}\alpha_{2,s}}
(16 \pi) \left(\frac{8}{9}\right)
\left(\frac{\hat u}{\hat t} + \frac{\hat t}{\hat u}\right)
\; , \\
\overline{|\mathcal{M}_{\bar q q\to \gamma X}|^2}
&=&
\pi \alpha_{em}\sqrt{\alpha_{1,s}\alpha_{2,s}}
(16 \pi) \left(\frac{8}{9}\right)
\left(\frac{\hat t}{\hat u} + \frac{\hat u}{\hat t}\right)
\; , \\
\overline{|\mathcal{M}_{g q\to \gamma X}|^2}
&=&
\pi \alpha_{em}\sqrt{\alpha_{1,s}\alpha_{2,s}}
(16 \pi) \left(-\frac{1}{3}\right)
\left(\frac{\hat u}{\hat s} + \frac{\hat s}{\hat u}\right)
\; , \\
\overline{|\mathcal{M}_{q g\to \gamma X}|^2}
&=&
\pi \alpha_{em}\sqrt{\alpha_{1,s}\alpha_{2,s}}
(16 \pi) \left(-\frac{1}{3}\right)
\left(\frac{\hat t}{\hat s} + \frac{\hat s}{\hat t}\right)
\; . 
\end{eqnarray*}
\end{widetext}
Including finite mass effects for quarks/antiquarks and off-shell
effects for gluons \cite{LZ05a} the matrix element can be written as:
\begin{widetext}
\begin{eqnarray*}
\overline{|\mathcal{M}_{q \bar q \to \gamma X}|^2} 
&=&
-\frac{8(4\pi)^2\alpha_{em}\sqrt{\alpha_{1,s}\alpha_{2,s}}}
{9(\hat t - m_q^2)^2(\hat u - m_q^2)^2} F_{q \bar q}(k_{1,t},k_{2,t})
\; , \\
\overline{|\mathcal{M}_{\bar qq \to \gamma X}|^2}
&=&
-\frac{8(4\pi)^2\alpha_{em}\sqrt{\alpha_{1,s}\alpha_{2,s}}}
{9(\hat u - m_q^2)^2(\hat t - m_q^2)^2} F_{\bar qq}(k_{1,t},k_{2,t})
\; , \\
\overline{|\mathcal{M}_{gq \to \gamma X}|^2}
&=&
\frac{(4\pi)^2\alpha_{em}\sqrt{\alpha_{1,s}\alpha_{2,s}}}
{3(\hat s - m_q^2)^2(\hat u - m_q^2)^2} F_{gq}(k_{1,t},k_{2,t})
\; , \\
\overline{|\mathcal{M}_{qg \to \gamma X}|^2}
&=&
\frac{(4\pi)^2\alpha_{em}\sqrt{\alpha_{1,s}\alpha_{2,s}}}
{3(\hat s - m_q^2)^2(\hat t - m_q^2)^2} F_{qg}(k_{1,t},k_{2,t})
\; , 
\label{LZ_MEs}
\end{eqnarray*}
\end{widetext}
where for brevity we have introduced
\begin{widetext}
\begin{eqnarray*}
F_{q\bar q}(k_{1,t},k_{2,t})
&=&
6m_q^8-(3\hat t^2 +3\hat u^2 +14\hat t\hat u)m_q^4 \nonumber \\
&+&(\hat t^3 + \hat u^3 +7\hat t \hat u^2 +7\hat t^2 \hat u^2)m_q^2
\nonumber \\
&-&\hat t \hat u(\hat t^2 + \hat u^2)
\; , \\
F_{\bar qq}(k_{1,t},k_{2,t})
&=&
6m_q^8-(3\hat u^2 +3\hat t^2 +14\hat u\hat t)m_q^4 \nonumber \\
&+&(\hat u^3 + \hat t^3 +7\hat u \hat t^2 +7\hat u^2 \hat t^2)m_q^2
\nonumber \\
&-&\hat u \hat t(\hat u^2 + \hat t^2)
\; , \\
F_{gq}(k_{1,t},k_{2,t})
&=&
6m_q^8-(2k_{1,t}^4+2(\hat s \hat u)k_{1,t}^2+
3\hat s^2+3\hat u^2+14\hat s \hat u)m_q^4 \nonumber \\
&+&(2(\hat s \hat u)k_{1,t}^4+8\hat s \hat u k_{1,t}^2
+\hat s^3+\hat u^3+7\hat s \hat u^2+7\hat s^2 \hat u)m_q^2 \nonumber \\
&-&\hat s \hat u(2k_{1,t}^4+2(\hat s \hat u)k_{1,t}^2
+\hat s^2 + \hat u^2)
\; , \\
F_{qg}(k_{1,t},k_{2,t})
&=&
6m_q^8-(2k_{2,t}^4+2(\hat s \hat t)k_{2,t}^2+
3\hat s^2+3\hat t^2+14\hat s \hat t)m_q^4 \nonumber \\
&+&(2(\hat s \hat t)k_{2,t}^4+8\hat s \hat t k_{2,t}^2
+\hat s^3+\hat t^3+7\hat s \hat t^2+7\hat s^2 \hat t)m_q^2
\nonumber \\
&-&\hat s \hat t(2k_{2,t}^4+2(\hat s \hat t)k_{2,t}^2
+\hat s^2 + \hat t^2)
\; . 
\end{eqnarray*}
\end{widetext}
In the formula above only transverse momenta 
of the ingoing gluons are included explicitly when calculating
matrix elements.
Usually gluons generated via QCD effects have on average larger
transverse momenta than quarks.



\end{document}